\documentclass[prb,twocolumn,showpacs,floatfix]{revtex4-1}
\usepackage{epsfig,amssymb,amscd}
\usepackage{times,amsmath,graphicx,amsfonts}

\newcommand{\non}{\nonumber}

\newcommand{\bea}{\begin{eqnarray}}
\newcommand{\eea}{\end{eqnarray}}
\newcommand{\be}{\begin{equation}}
\newcommand{\ee}{\end{equation}}
\newcommand{\ba}{\begin{align}}
\newcommand{\ea}{\end{align}}

\newcommand{\ket}[1]{     |    \,    #1    \rangle}
\newcommand{\bra}[1]{  \langle #1  \,  |} 
\newcommand{\ZZ}{\mathbb{Z}}

\newcommand{\crc}{c^\dagger}

\newcommand{\bk}{{\boldsymbol{k}}}

\newcommand{\bq}{{\boldsymbol{q}}}

\newcommand{\bs}[1]{\boldsymbol{#1}}

\begin{document}

\title{Topological Blocking in Quantum Quench Dynamics}

\author{G.~Kells$^{1,2}$, D. ~Sen$^3$, J.~K.~Slingerland$^{1,4}$, 
S.~Vishveshwara$^5$}

\affiliation{
$^1$Department of Mathematical Physics, National University of Ireland, 
Maynooth, Ireland \\
$^2$  Dahlem Center for Complex Quantum Systems and Fachbereich Physik, Freie Universit\"{a}t Berlin, Arnimallee 14, 14195 Berlin, Germany.\\
$^3$Centre for High Energy Physics, Indian Institute of Science, Bangalore 
560 012, India \\
$^4$Dublin Institute for Advanced Studies, School of Theoretical Physics, 
10 Burlington Rd, Dublin, Ireland \\
$^5$Department of Physics, University of Illinois at Urbana-Champaign, 
Urbana, Illinois 61801-3080, USA}

\begin{abstract}
We study the non-equilibrium dynamics of quenching through a quantum
critical point in topological systems, focusing on one of their defining
features--- ground state degeneracies and associated topological sectors. We 
present the notion of ``topological blocking", experienced by the dynamics due
to a mismatch in degeneracies between two phases and we argue
that the dynamic evolution of the quench depends strongly on the topological 
sector being probed. We demonstrate this interplay between quench and topology 
in models stemming from two extensively studied systems, the transverse Ising 
chain and the Kitaev honeycomb model. Through non-local maps of each of these 
systems, we effectively study spinless fermionic $p$-wave paired superconductors. 
Confining the systems to  ring and toroidal geometries, respectively, enables us to cleanly address degeneracies, subtle issues of 
fermion occupation and parity, and mismatches between topological sectors.  We show that various 
features of the quench, which are related to Kibble-Zurek physics, are 
sensitive to the topological sector being probed, in particular, the overlap 
between the time-evolved initial ground state and an appropriate low-energy 
state of the final Hamiltonian. While most of our study is confined to 
translationally invariant systems, where momentum is a convenient quantum 
number, we briefly consider the effect of disorder and illustrate how this can influence the quench in a qualitatively different way depending on the topological sector considered. 
\end{abstract}


\date{\today}
\maketitle
\section{Introduction}
Over the past years, there has been a revival of interest in the topics of topological 
systems and non-equilibrium critical dynamics stemming from the latest 
advances exhibited in a variety of condensed matter and cold atomic systems
~\cite{Dziarmaga10,Kibble76,Zurek96,Damski05,Anatoly05,Calabrese05,Levitov06,
Mukherjee07,Sengupta08,Mondal08,Patane08,Degrandi08,Bermudez09,Perk09,Polkov08, 
Vishveshwara10,Pollmann10,Singh10,Sondhi12,Sondhi13,Patel13,Mostame13,
Foster13}. The synergy of the two topics, namely quench dynamics 
in topological systems, is still in  its infancy~\cite{Bermudez09,Vishveshwara10,Sondhi12,Sondhi13,Patel13}, 
but promises to form a rich and complex avenue of study.  While previous works have targeted the formation
 of edge states and bulk defects that are characteristic of topological systems, in this work we  focus in
 particular on the role of ground-state degeneracy, another key characteristic of topological order.

Our work highlights special features of quenches that involve initializing a 
system in the ground state of a phase with a particular topological order and 
dynamically evolving this state through a topological phase transition, i.e. the Hamiltonian is time dependent and the ground-states of the initial and final Hamiltonians have differing topological order. We consider topological aspects of 
systems having periodic boundary conditions, i.e., rings or tori, where the 
effect of degeneracies is clear cut. This is different from open bounded systems, where the dynamics can be complicated by edge effects and from infinite systems where topological aspects can often be completely hidden.  Most dramatically, we find a phenomenon which we call \emph{topological blocking}: due to mismatch in degeneracies, some of the ground states of a 
topological system have no overlap with any of the ground states on the other 
side of the transition, regardless of how slowly the quench is performed.

\begin{figure}
\includegraphics[width=0.4\textwidth,height=0.22\textwidth]{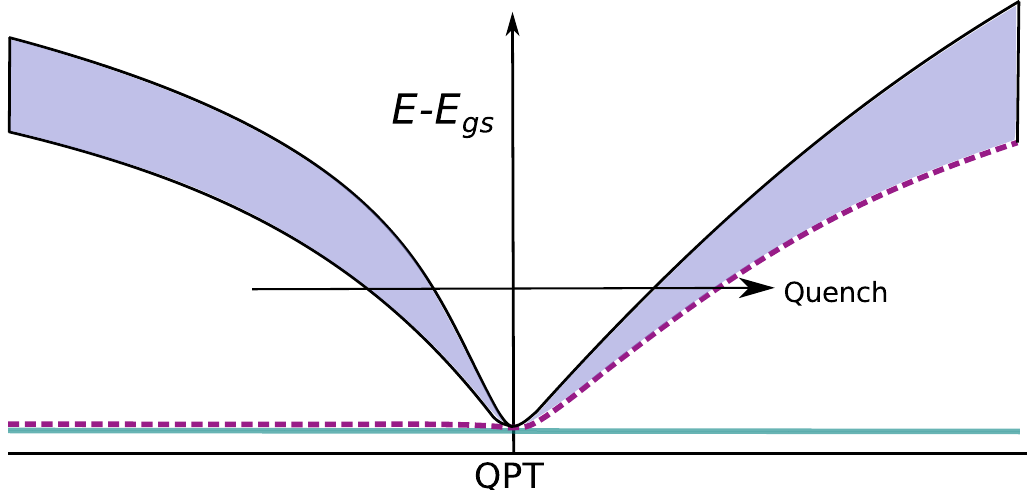}
\caption{(Color online) An example of topological blocking in which the quench goes from a 
doubly degenerate gapped  phase to a non-degenerate gapped phase, as happens, for instance, in the 
one-dimensional spinless $p$-wave superconductor. Upon closing the gap at the critical point, one of the degenerate
 states in the initial phase is lifted into the continuum of excited states in the final phase.
This time-evolved initial state has no overlap with the final ground state.} 
\label{fig:TopQuench} \end{figure}

We expect that our central observations apply to a wide range of topological 
systems. Our general setting involves two gapped phases having different
degeneracies separated by a gapless critical point (or more generally, a 
gapless region). Topological blocking is best seen by initializing the system 
in the phase having higher degeneracy. Over the evolution of the quench, as 
shown in  Figure~\ref{fig:TopQuench} some of the topological sectors of 
this phase are forced to be lifted in energy as they pass through the gapless 
point so that no states in those sectors appear as ground states in the new 
phase. Nevertheless the states in the original topological sectors may remain 
topologically distinct from each other, so they cannot be connected by the 
action of local operators. Hence, in a quantum quench between the phases, an 
initial state in a sector that has its energy lifted evolves within that 
sector. The time evolved state after the quench thus has zero overlap with 
any of the ground states in the final phase.

\begin{figure}
\includegraphics[width=0.45\textwidth,height=0.36\textwidth]{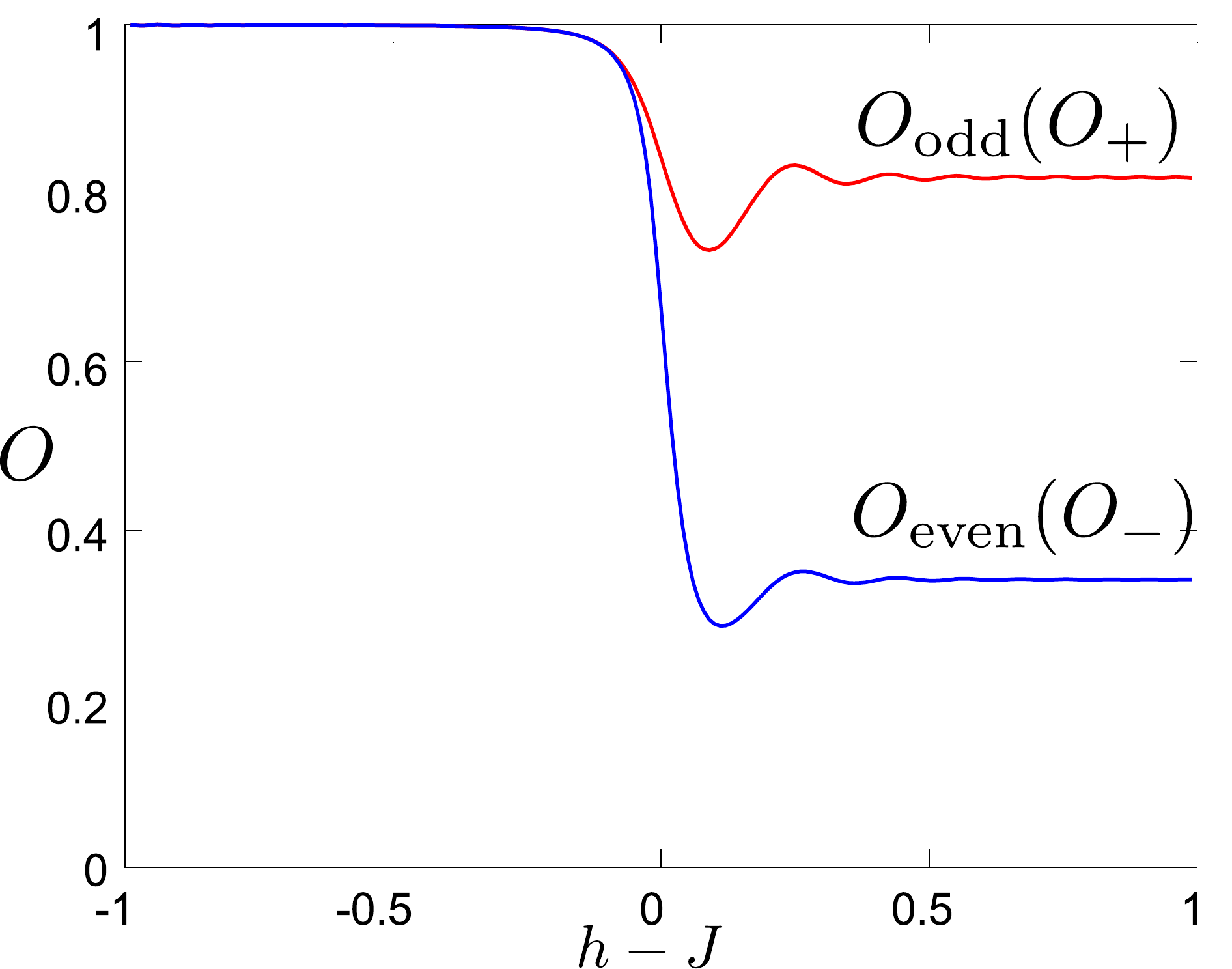}
\caption{(Color online) Typical  quench data for the one-dimensional spinless $p$-wave superconductor
 on a $N=60$ site ring shows the overlap of the time-evolved ground state of the  
initial Hamiltonian in one topological sector with the lowest energy state of the final Hamiltonian
 within that sector. (This is not always the global ground state of the final Hamiltonian.) As a result of the topological blocking, the odd-fermion sector with 
periodic boundary conditions has a higher final overlap than the even-fermion sector.} 
\label{fig:TrIsing_example} \end{figure}

The role of the topological sector, while directly obvious for topological 
blocking, is also apparent when considering state evolution within the sector. 
We find that an effective indicator of sectoral-dependence is the overlap of the
time-evolved  state with the lowest energy state of  the instantaneous quenched Hamiltonian within the same sector (sectoral 
ground state). Figure~\ref{fig:TrIsing_example} shows an example illustrating 
such time-dependent wave function overlaps for a quench from a doubly 
degenerate phase to a non-degenerate phase; the overlaps within the two 
sectors, labeled by parity, show a clear difference in their evolution during 
the quench, exhibiting the most pronounced features in the vicinity of the
critical point. While the quantitative difference is obvious, under certain 
easily accessible circumstances, there can also be a qualitative difference 
if, unlike the absolute ground state, some of the sectoral ground states 
in the post-quench phase are not separated by a gap from the spectrum of 
excited states. It is worth mentioning here that these systems still respect 
the well-studied Kibble-Zurek mechanism~\cite{Kibble76,Zurek96,Damski05,Anatoly05,Levitov06,Mukherjee07}, 
which applies to systems having local as well as topological order and 
predicts power-law scaling as a function of quench rate in various quantities
related to post-quench excitations. The dependence on topological sectors 
rides above such scaling and among the typical Kibble-Zurek quantities, such 
as residual energy or defect density, is most strongly manifest in 
wave function overlaps.

 The interplay between topology and quench dynamics provides new insights into
each of these respective aspects. Our treatment shows that the quench dynamics 
between phases that have different ground state degeneracies acts as a fine probe 
of topological order and examines some of its more subtle issues. For example, 
the notion of topological blocking highlights the fact that the number of 
topologically distinct subspaces (sectors) of the Hilbert space of a system 
may exceed the ground state degeneracy; there may be topological sectors 
which are ``hidden" at low energy, but which nevertheless play a role in 
quantum quenches. In terms of quench physics, we bring attention to the 
concept that there typically exist multiple sectors in a system having 
topological order, which could show distinctly different dynamics. 
Understanding these quenches is also essential for the implementation of 
topologically fault tolerant quantum computation 
schemes~\cite{Nayak08,Alicea11,Vishveshwara11} where collective transitions 
between topological and non-topological phases (see for example 
Ref.~\onlinecite{Kells2013}) represent a potential source of decoherence. 
The topological blocking mechanism and the fact that the ``hidden" 
topological sectors need not be gapped (as we show below), presents a 
further complication for such schemes.

In what follows, we perform an analysis of the features we discussed above 
within the context of two topological systems that can effectively be 
described as spinless fermionic $p$-wave paired superconductors. We first 
study the quantum Ising chain in a transverse magnetic field, perhaps one of 
the most celebrated systems in condensed matter for offering a tractable 
solution and rich physics, one with plentiful studies even in the context of 
quenching~\cite{Levitov06,Mukherjee07,Calabrese2012,Essler2012,Fagotti13,Barouch1970}.
The second system, the Kitaev honeycomb model, too is special in its 
analytically soluble 
structure~\cite{Kitaev2006,Chen2007,Baskaran07,Kells2008,Kells2009,Schmidt2007,
Kells2010,Knolle13} and has also received significant attention in the context
of quenching~\cite{Sengupta08}. The transverse Ising model maps to a $p$-wave 
superconducting chain~\cite{Kitaev2001} while the honeycomb lattice model 
maps to a $p+ip$ superconductor coupled to a $\ZZ_2$ gauge 
field~\cite{Kitaev2006,Chen2007,Kells2008,Kells2009}, the latter thus a 
natural two-dimensional extension of the former.

In these superconducting systems, topological sectors are identified in 
terms of fermion parity, which is naturally accounted for in the ring and 
torus topologies for the one- and two-dimensional cases, respectively. In 
the transverse-Ising systems, the quench involves going from a topological phase having 
double degeneracy associated with even and odd parity to a non-topological 
phase having a unique ground state characterized by one of the two 
parities. In the spin language, the phase with the two degenerate ground 
states corresponds to a ferromagnet that spontaneously breaks local Ising 
$\ZZ_2$ symmetry while that with the unique ground state corresponds to a 
phase with spin-polarization along the magnetic field. In the honeycomb lattice model 
the relevant phases are an Abelian phase having the topological order of 
the toric code, which is four-fold degenerate, and a non-Abelian phase with 
Ising type topological order and three-fold degeneracy. In this model the 
transition is topological both in the spin language and in the fermionic 
language. In both models, we carefully pinpoint how topological blocking 
comes about, using the structure of the Bogoliubov-deGennes (BdG) 
Hamiltonians and perform a detailed analysis of the difference in post-quench 
behavior for quenches within different topological sectors. 

The mapping in the transverse-Ising system between a model having local $\ZZ_2$ 
symmetry and one with topological order begs for a comment on the relevance 
of our analyses to systems having spontaneous symmetry breaking and local 
order. As with topological systems, in quenching through a spontaneous 
symmetry breaking transition, the symmetry broken phase would typically 
have larger ground state degeneracy than the unbroken phase and, if the 
quench dynamics preserves the symmetry, a similar blocking phenomenon can 
occur; some symmetry breaking states would be lifted away from the ground 
state energy in the unbroken phase. In fact, much of our analysis
would apply for these systems and it would be worth studying sectoral 
dependences in the context of local order as well. However, an important 
distinction of topological blocking is the non-local 
nature of topological symmetries. Thus, unlike in spontaneous symmetry 
broken systems, the key features of topological blocking discussed in 
this work should be robust against local perturbations of the 
Hamiltonian of the system. 

An overview of the paper is as follows.
Section~\ref{sec:Ising} discusses the transverse-Ising case in depth, 
starting with a brief introduction, followed by its superconductor 
description, a discussion of degeneracies and the quench protocol, an 
explanation of topological blocking in terms of parity arguments, and finally 
detailed studies of quench behavior for different topological sectors. 
Section~\ref{sec:Honeycomb} gives a similar treatment of the Kitaev honeycomb 
model. In 
Sec.~\ref{sec:disorder}, we perform initial studies of quenches in these
systems in the presence of disorder as a means of demonstrating robustness 
against local perturbations as well as the marked difference in topological 
sectors in situations where the blocked sector can access a slew of low-lying 
excitations. We conclude with a short summary and outlook in 
Sec.~\ref{sec:SnO}.

\section{The transverse Ising model}
\label{sec:Ising}

The transverse Ising model in one dimension is one of the best studied 
exactly solvable models, (see Ref.~\onlinecite{Sachdev_book} for a thorough treatment). As is commonly done to solve 
almost any aspect of the model, the non-local Jordan-Wigner transformation 
is used to map it to a beautiful prototype of a topological system - a spinless
fermionic, one-dimensional $p$-wave superconductor. Here, after introducing 
the model, we reiterate the fermionization procedure, taking into account the
 subtleties associated with periodic boundary conditions and fermion parity. 
We carefully describe the link between fermion parity, topological degeneracy,
 the topological sectors on either side of the transitions and their associated 
sectoral ground states. With these considerations in place, we show how 
topological blocking naturally comes about. We then study the dynamics of the 
quench in each topological sector, focusing on the overlap between the 
time-evolved initial ground state and instantaneous sectoral ground states. 
Our analytic treatment uses the Landau-Zener formalism typically 
applied of late to related quenches in homogeneous 
systems~\cite{Damski05,Anatoly05,Levitov06,Mukherjee07} and we 
corroborate it with numerical studies. 

The most frequently encountered form of the Hamiltonian for the transverse 
Ising model is given by
\be H_{TI}= -J \sum_{<ij>} \sigma^x_i \sigma^x_j - h \sum_i \sigma_i^z. 
\label{ham1} \ee
Here, $\sigma^i$ denote spin $1/2$ Pauli matrices, $J$ an Ising ferromagnetic 
coupling, $h$ a Zeeman magnetic field in the $z$-direction, and $<ij>$ 
nearest neighbors $i$ and $j$. (We set Planck's constant $\hbar = 1$ 
throughout this paper). If we take $J>0$ and $h>0$, the system has two phases, ferromagnetic and 
paramagnetic. The ordered Ising ferromagnet along 
the $x$-direction occurs for $h < J$ while the paramagnetic phase occurs for 
$h > J$. The two phases are separated by a quantum critical point at $h=J$.

The ground state degeneracies of the two phases can be discerned by looking 
at the Hamiltonian in some simple limits. In the paramagnetic limit, $J=0$, 
we see that the ground state is simply the non-degenerate state fully 
polarized along the direction of the Zeeman magnetic term,
\be \ket{\text{gs}} = \ket{\bar{0}}=\ket{00...00}, \label{gs} \ee
where, for the spin state on a single site, $\ket{0} = [1,0]^T$ and $\ket{1} = [0,1]^T$
 in the eigenvalue basis of $\sigma^z$. The overbar denotes the quantum state for 
the entire collection of sites. In the opposite ferromagnetic limit, $h=0$, there are two
degenerate ground states given by superpositions of
\be \ket{\bar{+}}= \ket{++...++} \;\;\; \text{and} \;\;\; \ket{\bar{-}}= 
\ket{--...--}, \label{twogs} \ee
where $\ket{+} = [1,1]^T/\sqrt{2}$ and $\ket{-} = [1,-1]^T\sqrt{2}$ are the 
eigenstates of $\sigma^{x}$. The system is symmetric under a global $\pi$ 
rotation around the $z$-axis, given (up to a global phase) by the string 
operator
\be T_z=\prod_i \sigma^z. \label{eq:string} \ee
This non-local operator maps the $\ket{\bar{+}}$ and $\ket{\bar{-}}$ states 
into each other, while $\ket{\bar{0}}$ is left invariant. After 
fermionization, $T_z$ is associated with fermion parity and topological 
degeneracy; note that $T_z$ is conserved even if the couplings in 
Eq.~\eqref{ham1} are allowed to be functions of space. 

\subsection{Fermionized topological superconductor and solution}
\label{sec:IsingFermionic} 

The original fermionic solution for the transverse Ising chain can be 
traced to Pfeuty~\cite{Pfeuty1970} who used a transformation similar to 
Lieb, Schultz and Mattis~\cite{Lieb1961}. Indeed, 
the fermionic dispersion relation for the transverse Ising can be seen to be 
identical to that of the $XY$ model solved by Lieb, Schultz and Mattis.
Here too we employ their extensively used Jordan-Wigner transformations 
to define the position space fermionic excitations (see, for example, 
Refs.~\onlinecite{Levitov06,Mukherjee07,Calabrese2012}) 
\be c^\dagger_i = ( \prod_{j<i} \sigma_j^z ) ~\sigma_i^- ~~~{\rm and}~~~
c_i = ( \prod_{j<i} \sigma_j^z ) ~\sigma_i^+. \ee 
The state $\ket{\bar{0}}$ given in Eq.~\eqref{gs} is therefore the fermionic 
vacuum state. At any site $i$, we have $\sigma_i^z = (-1)^{c_i^\dagger c_i}$. 
Hence $T_z$ gives the parity of the total fermion number,
\be T_z = (-1)^{N_F} ~~\mathrm{with}~~N_F =\sum_i c_i^\dagger c_i. \ee
In terms of fermion operators the Hamiltonian takes the superconducting form
\bea H &=& h \sum_{i=1}^N (2c^\dagger_i c^{\phantom \dagger}_i - 1) \non \\
\non &&- J \sum_{i=1}^{N-1} (c^\dagger_i-c^{\phantom \dagger}_i)
(c_{i+1}^\dagger +c_{i+1}^{\phantom \dagger}) \non \\
&&+ J T_z (c^\dagger_N-c^{\phantom \dagger}_N)(c_1^\dagger+
c_1^{\phantom \dagger}), \label{eq:HamBdg} \eea
where $N$ is the number of sites on the ring. This superconducting Hamiltonian 
for spinless fermions has an on-site chemical potential $\mu= -2h$, 
nearest-neighbor hopping of strength $w=J$, and anomalous $p$-wave pairing 
terms also of strength $\Delta = J$. A generalization of this model having 
$w\neq \Delta$ can be obtained by considering an $XY$ spin chain instead of 
an Ising spin chain~\cite{Mukherjee07}; 
the main results of this section also hold for this case.

The boundary conditions of the system are encoded in the operator $T_z$. 
To select the periodic sector we replace the operator $T_z$ with its 
eigenvalue $-1$ corresponding to an odd number of fermions. To select the 
antiperiodic sector we replace the operator with the eigenvalue $+1$,
corresponding to even parity.

The Hamiltonian can be written in momentum space as a sum of BdG Hamiltonians
\bea H &=& \sum_{0 \le k \le \pi} \left[\begin{array}{cc} 
c^\dagger_{k} & c_{-k}
\end{array} \right] H_k \left[\begin{array}{c}
c_{k}
\\ c^\dagger_{-k} \end{array} \right], \non \\
{\rm where} ~~~~H_k &=& \left[\begin{array}{cc} 
\xi_k & \Delta_k \\ \Delta_k^* & -\xi_k \end{array} \right], \non \\
\xi_k &=& -\mu -2 w \cos (k), \non \\
\Delta_k &=& 2 w \sin(k). \eea
The BdG Hamiltonians $H_k$ can be diagonalized by a Bogoliubov transformation.
Namely, we may write
\bea H &=& \sum_{0 \le k \le \pi} \epsilon_k (\gamma^{\dagger}_{k} 
\gamma^{\phantom \dagger}_k + \gamma^{\dagger}_{-k} 
\gamma^{\phantom \dagger}_{-k} -1), \label{eq:Hdiag} \non \\
\epsilon_k &=& \sqrt{\xi_k^2 +|\Delta_k|^2}, \label{xidelta} \eea
in terms of the Bogoliubov-Valatin operators
\bea \label{eq:gamma1}
\gamma_{k} &=& u_k c^{\phantom \dagger}_k - v_k c^{\dagger}_{-k}, ~~~~
\gamma_{k}^\dagger = u_k^* c^\dagger_k - v_k^* c^{\phantom \dagger}_{-k}, 
\non \\
\gamma_{-k} &=& u_k c^{\phantom \dagger}_{-k} + v_k c^{\dagger}_{k}, ~~~~
\gamma_{-k}^\dagger = u_k^* c^\dagger_{-k} + v_k^* c^{\phantom \dagger}_{k}, 
\eea
with
\bea \label{eq:uv1}
u_k &=& \phantom -\sqrt{(1 + \xi_k/\epsilon_k)/2}, \non \\
v_k &=& -\sqrt{(1 - \xi_k/ \epsilon_k)/2}. \eea
(We will see below that the modes with $k = 0$ and $\pi$ require a special
analysis since they satisfy $k = - k$. Further, $\Delta_k = 0$ for these 
modes; hence, $\varepsilon_k = |\xi_k|$.)
We see that in both phases of the model, the excitation energy $\epsilon_k$ 
is gapped for all $k$; the minimum energy lies at $k = 0$ with $\epsilon_0 
= 2 |h-J|$. At the critical point $h = J$, the system
is gapless and $\epsilon_k = 0$ for $k = 0$.

With regard to the topological aspects of the superconductor, 
the ferromagnetic
phase, having a double ground state degeneracy, maps to a topological phase and
the non-degenerate paramagnetic phase to a topologically trivial phase. This 
can be seen from standard Berry's phase analyses of the momentum eigenstate 
spinor structure~\cite{Niu2012}. Alternatively, it is common
to consider the Kitaev chain, a finite open chain version of the Hamiltonian in 
Eq.~\eqref{eq:HamBdg}, which naturally lacks the $T_z$ term associated with
the (anti)periodic boundary conditions of the ring geometry. The topological 
phase then has free Majorana modes at each end which lie at zero energy if
the chain length is much larger than the decay length of these end modes.
The Majorana end modes together form a Dirac 
fermion state which can either be occupied or unoccupied, thus accounting for 
the double degeneracy and fermion parity. As alluded to above and detailed in 
what follows, for the ring geometry, which we confine ourselves to, the 
connection between topological degeneracy and fermion parity is more subtle. 

\subsection{Topological degeneracy}

\begin{figure}
\includegraphics[width=0.45\textwidth,height=0.25\textwidth]{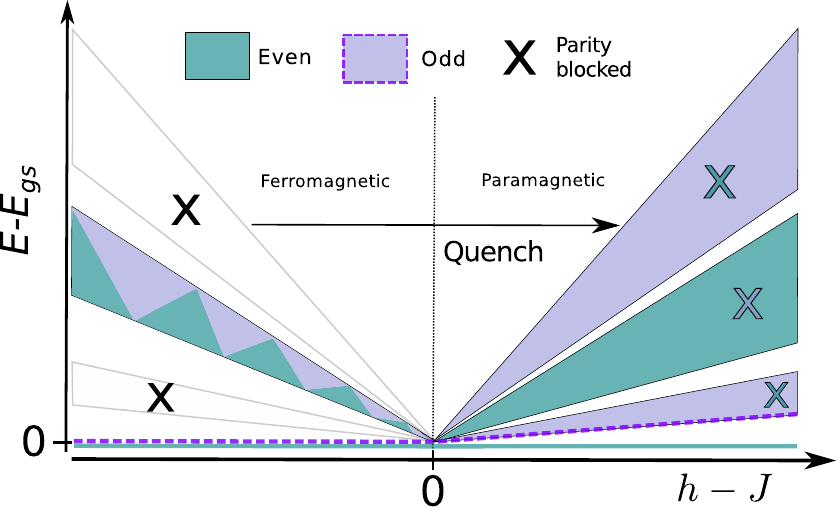}
\hspace*{.02cm} 
\caption{(Color online) Schematic of the spectrum of the transverse Ising ring as a 
function of $h-J$. In the ferromagnetic phase, the ground state is doubly 
degenerate in the thermodynamic limit and the excitation spectrum consists 
of bands of states with both even and odd fermion numbers. These states are 
created from the two ground states using pairs of $\gamma^{\dagger}$ 
operators. In particular, there are no energy levels with an odd number of 
$\gamma^{\dagger}$ excitations over one of the ground states. We explicitly 
indicate these levels as the parity blocked regions. In the paramagnetic phase
 there is a unique ground state with an even number of fermions. The lowest 
excited band consists of odd fermion number states which are however not 
created by single $\gamma^{\dagger}$-s from the ground state. Further bands 
are created from ground state and the lowest band using pairs of 
$\gamma^{\dagger}$ operators. The purple dashed line indicates that, in the adiabatic limit, the odd sectoral ground-state of the ferromagnetic phase flows to the lowest energy state in the paramagnetic phase.}
 \label{fig:isingchainspec} \end{figure}

We now describe the ground states of the model in terms of the occupation 
numbers of the fermionic modes and explain in detail how the topological 
sectors of the Ising chain are connected to fermion parity. In particular, we show
 that there is always a ground state of the system with even fermion number, 
while a ground state with odd fermion number exists only in the ferromagnetic
 phase. In the paramagnetic phase, the lowest energy state with odd fermion
 number is part of a band which is gapped away from the true (even fermion 
number) ground state. A schematic of the spectrum of the model highlighting 
these features is shown in Fig.~\ref{fig:isingchainspec}. 

We focus first on the case where the number of sites $N$ is even.
In the even-fermion antiperiodic sector, the allowed momenta are then 
given by $k = \frac{2 \pi}{N} (n+\frac{1}{2})$ with integer $n \in 
[-N/2,N/2-1]$. Crucially, note that the values of $k$ do not include $0$ and 
$\pi$. The ground state is given by
\be \ket{\text{gs}}_{\text{even}}= \prod_{0 < k < \pi,~\frac{Nk}{\pi}~\mathrm{odd}} (u_k + v_k 
\crc_k \crc_{-k} ) \ket{\bar{0}}, 
\label{gseven} 
\ee
where $k$ spans the restricted set of momenta described above. The energy
 of this state is given by$E_{\text{gs}} =-\frac{1}{2} \sum 
\epsilon_k $, where the sum respects the quantization condition on $k$.

In the odd-fermion periodic sector the allowed momenta are given by $k = 
\frac{2 \pi n}{N}$ with integer $n \in [-N/2,N/2-1]$. These include the 
momenta $k=0,\pi$, which need to be treated carefully. In the ferromagnetic 
phase occurring for $J > h \ge 0$, we have $u_0=0,v_0=1$; hence 
$\gamma_0=c^\dagger_0$. From Eq.~\eqref{eq:Hdiag} we see that the 
contribution of this mode to the Hamiltonian is then just $H_0=-
2(h-J)(c_0^\dagger c_0^{\phantom \dagger} - 1/2)$, and thus the fermionic 
state with the $k=0$ mode occupied has the lower energy compared to that
with the mode unoccupied. We also have $u_\pi = 1, v_\pi = 0$, so that 
$\gamma_\pi = c_\pi$, and similar arguments show that the energetically 
favorable state has the $k=\pi$ mode unoccupied. Hence the ground state is 
given by
\be 
\ket{\text{gs}}_{\text{odd}} = c^\dagger_0 \prod_{0 < k < \pi,~\frac{Nk}{\pi}~\mathrm{even}} (u_{k} 
+ v_{k} \crc_{k} \crc_{-k} )\ket{\bar{0}}, \label{gsodd1} 
\ee
As this  state is annihilated by all the $\gamma_k$ it has an energy given
by $E_{\text{gs}}=-\frac{1}{2} \sum \epsilon_k$. In this phase, the values 
of $k$ become arbitrarily close to those of the even-fermion sector and for 
$N \gg 1$ we get a two-fold degenerate ground state.

To get an intuitive picture of how the degeneracy arguments derived from 
parity considerations connect with the spin picture described earlier, 
we can analyze the limit $h=0$. 
For any value of $N$, we then have two degenerate ground states given by all 
$\sigma_i^x = +1$ or all $\sigma_i^x = -1$ as shown in Eq.~\eqref{twogs}. In 
terms of states with fermionic occupation numbers $\ket{0}_i$ and $\ket{1}_i$ 
at site $i$, the two ground states are given by $\ket{\bar{+}} = \prod_i 
(\ket{0}_i + \ket{1}_i)/\sqrt{2}$ and $\ket{\bar{-}} = \prod_i (\ket{0}_i - 
\ket{1}_i)/ \sqrt{2}$. We then see that the sum and difference of these states 
respectively give states which have an even and odd number of fermions, 
recalling that $N$ is even.

The situation is quite different in the paramagnetic phase which 
occurs for $h > J \ge 0$. The odd-fermion parity sector has a state
with $k = 0$ with $u_0=1,v_0=0$, so that $\gamma_0 = c_0$. In principle, 
having the fermionic $k=0$ and $k=\pi$ modes unoccupied would be the lower
energy state. However, this would violate the odd parity of the sector. 
Given that as a function of $k$, $\epsilon_k$ has the smallest value for 
$k = 0$, the state defined in Eq.~\eqref{gsodd1} still does the best in 
terms of minimizing the energy within the odd sector. In this case, 
$c_{0}^{\dagger}=\gamma_{0}^{\dagger}$, so we are looking at the state in 
Eq.~\eqref{gseven} with an extra $\gamma_{0}^{\dagger}$ excitation. This 
state is the lowest state of a band which can be obtained by exciting the 
system at nonzero momentum using $\gamma_{k}^{\dagger}$ instead of 
$\gamma_{0}^{\dagger}$. Thus, Eq.~\eqref{gsodd1} corresponds to the 
\textit{sectoral ground state} in the paramagnetic phase. 
However, the state now possesses energy $E_{\text{gs}} = \epsilon_0-
\frac{1}{2} \sum_k \epsilon_k$. In the limit $N \gg 1$,
we see that the ground state in the odd-fermion sector lies 
at an energy which is higher than the ground state in the even-fermion parity 
sector by a finite amount equal to $\epsilon_0 = 2 (h-J)$.

Now let us briefly discuss what happens if $N$ is
odd. Then in the even-fermion antiperiodic sector, the allowed momenta are 
given by $k = \frac{2 \pi}{N}(n+\frac{1}{2})$ with integer $n \in 
[-(N-1)/2,(N-1)/2]$, which includes the $k=\pi$ term but not $k = 0$. 
In both the ferromagnetic and paramagnetic phases, the even sectoral ground 
state is still given by Eq.~\eqref{gseven} (with the appropriate momentum 
quantization) and this state continues to be the absolute ground state.
In the odd-fermion periodic sector, the allowed momenta are given by $k 
= \frac{2 \pi n}{N}$ with integer $n \in [-(N-1)/2,(N-1)/2]$, which includes 
the $k=0$ term but not $k = \pi$. Here too, Eq.~\eqref{gsodd1} remains
the odd sectoral ground state and is another absolute ground state in the
ferromagnetic phase but has higher energy in the paramagnetic phase. 
The situation is therefore similar in many ways to the case where $N$ is even.

To summarize, in the thermodynamic limit $N \gg 1$, the 
ground state of the system in the ferromagnetic phase has a double degeneracy,
with one ground state lying in each of the sectors (even- and odd-fermion). 
In the paramagnetic phase, there is a unique ground state which lies in the 
even-fermion sector. The sectoral ground state in the odd-fermion sector lies 
in a band which is 
separated by a finite gap from the ground state in the even-fermion sector.

\subsection{Quenching Dynamics}

We now turn to the quench dynamics caused by slowly varying the 
transverse field in time, 
starting at $t=0$ at $h_i = 0$ in the ground state of the ferromagnetic phase 
and ending at $t=T$ at $h_f = 2J$ in the paramagnetic phase. Note that the 
time evolution does not mix the even- and odd-fermion sectors; hence we will 
consider the time evolution in the two sectors separately.

{\bf Quench protocol:-} We consider a linear time dependence of the form
\be h(t) = 2Jt/T, ~~ {\rm for} ~~ 0<t<T. \label{eq:hquench} \ee
 By a slow variation, we mean that the dimensionless quantity $JT \gg 1$.
Our analysis of quench dynamics partially follows those extensively performed
in the context of Kibble-Zurek physics~\cite{Damski05,Anatoly05,Levitov06,
Mukherjee07} with the crucial difference that we explicitly consider fermion 
parity and momentum quantization associated with the topological sectors. 

For any given set of $k$ modes (except 0 and $\pi$), the quench couples 
the two states in the occupation number basis $|n_k, n_{-k} \rangle =$
$|0,0\rangle$ and $|1,1\rangle$. In this basis, the relevant dynamics is 
governed by the Hamiltonian 
\be H_k (t) ~=~ J ~\left( \begin{array}{cc}
(t-a_k)/\tau & b_k \\
b_k & - (t-a_k)/\tau \end{array} \right), \label{hamlz1} \ee
where Eqs.~\eqref{xidelta} imply that 
\be \tau = \frac{T}{4}, ~~~~a_k = \frac{T}{2} \cos (k), ~~~~b_k
= 2\sin (k). \label{tauab1} \ee
The instantaneous eigenvalues of the Hamiltonian in Eq.~\eqref{hamlz1} have 
a minimum difference gap of $2b_k$ at $t=a_k$. In our problem, the value 
of $a_k$ depends on $k$. Further, the initial and final values of 
$t-a_k$ are given by 
\be t_{i,k} ~=~ -\frac{T}{2} \cos (k) ~~~~{\rm and}~~~~ t_{f,k} ~=~ T - 
\frac{T}{2} \cos (k) \label{tif} \ee
which also depend on $k$.

\begin{figure}
\includegraphics[width=0.50\textwidth,height=0.36\textwidth]{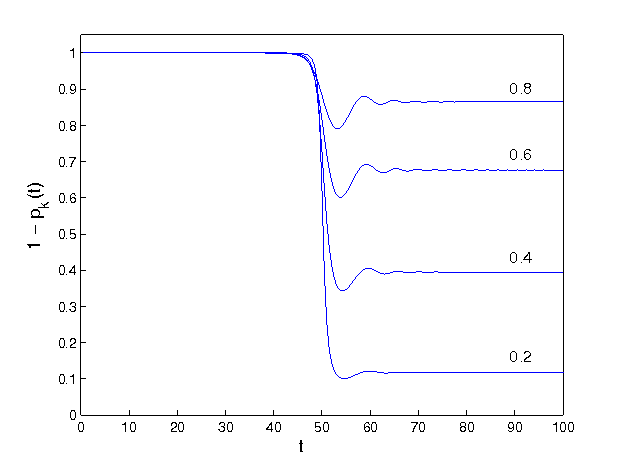}
\caption{(Color online) Overlap versus time for four two-level systems 
corresponding to $k \sqrt{JT} = 0.2$, $0.4$, $0.6$ and $0.8$, with $h(t) = 
2Jt/T$ and $JT=100$. We have set $J=1$. } 
\label{fig:overlap} \end{figure}

For each value of $k$, we study the quenching dynamics numerically as follows.
We first calculate the quantities $u_k (t)$ and $v_k (t)$ in 
Eqs.~(\ref{eq:gamma1}-\ref{eq:uv1}) at the initial time $t=0$ with the 
initial value $h=h_i$. We then compute the time ordered evolution operator
\be U_k (t,0) ~=~ {\cal T} ~[ \exp (-i \int_0^t H_k(t') dt')] \ee
by dividing the time $t$ into $N_t$ steps of size $\Delta_t$ each
(with $N_t \Delta_t = t$) and calculating
\be U_k (t,0) \approx \prod_{n=1}^{N_t} [ \exp( -i H_k(t_n) \Delta_t)], \ee
where $t_n = (n-1/2) \Delta_t$. We then calculate
\be \left( \begin{array}{c} 
u_k^* (t) \\ - v_k^*(t) \end{array}\right) ~=~ U_k (t,0) 
\left( \begin{array}{c} 
u_k^* (0) \\ - v_k^*(0) \end{array}\right). \label{eq:Ut} \ee
Finally we compute the ground state overlap by using the Onishi 
formula~\cite{Ring04} which, for our $2 \times 2$ matrices, amounts to 
\bea 
O_{\pm}(t) &=& | \langle gs  |\psi (t) \rangle|^2 \non \\
&=& \prod_{k}|\langle gs_k |\psi_k (t)\rangle|^2 \non  \\
&=& \prod_{k} | u_{k}^* (t) u_{k} + v_{k}^* (t) v_{k} |. 
\label{eq:Opm}
\eea
where the time independent quantities $v_k$ and $u_k$ are those given in 
Eq.~\eqref{eq:uv1} and encode the instantaneous ground state. Here, the subscript 
$\pm$ indicates the fermion parity, and consequently, the boundary conditions.
The product over $k$ runs over the entire Brillouin zone from $-\pi$ to $\pi$ and,
as discussed in previous sections, is restricted to certain values that depend on
fermion parity. For a given momentum pair,
the probability of being in the excited state of the Hamiltonian $H_k$ is 
\be p_k(t) = 1- |\langle gs_k |\psi_k (t)\rangle|^2. \label{pkt} \ee
This excitation probability governs much of the post-quench behavior. A plot 
of $1-p_k$ for a number of $k$-values can be seen in Fig.~\ref{fig:overlap}

{\bf Analysis:-} 
Because the fermion number parity is conserved throughout the quench, we 
observe the topological blocking behavior described in the introduction. 
Initializing the system in the ground state of the ferromagnetic/topological 
phase 
in the odd parity sector, we observe that, even at adiabatically slow quench 
rates, this state does not evolve to the overall ground state (which has even 
fermion number), but rather to the sectoral ground state in the odd fermion 
number band.

At non-adiabatic quench rates, we therefore consider the overlap of the 
time-evolved state with the sectoral ground state of the final Hamiltonian. 
Figure~\ref{fig:TrIsing_example} shows a representative case for the overlap 
as a function of time for the odd- and even-fermion sectors; the two curves are
clearly different. We now analyze the detailed behavior of the time-evolved 
states, focusing on the contributions of each of the momentum modes and on 
the differences between sectors. 

To begin with, we consider a 
simple problem in which the time $t$ in Eq.~\eqref{hamlz1} goes from $-\infty$ 
to $\infty$, so that the value of $a_k$ is irrelevant. If we start in 
the ground state of $H(t)$ at $t=-\infty$, the probability of ending in the 
excited state of $H(t)$ at $t=\infty$ is given by the Landau-Zener 
expression~\cite{Landau32,Majorana32,Vitanov96}
\be p_k (t=\infty) ~=~ e^{-\pi J b_k^2 \tau} ~=~ e^{-\pi J T \sin^2 (k)}.
\label{lz} \ee
This expression gives the correct limits $p_k (\infty) \to 0$ and 1 in the 
adiabatic ($T \to \infty$) and sudden ($T \to 0$) limits respectively. 
Note that the momenta $k = 0$ and $\pi$ are special; $b_k = 0$ for these
modes and therefore $p_k = 1$ for any quenching time $T$. Namely, these
states do not change at all under quenching, and they change abruptly from 
the ground state to the excited state when $t$ crosses zero.

\begin{figure}
\includegraphics[width=0.48\textwidth,height=0.18\textwidth]{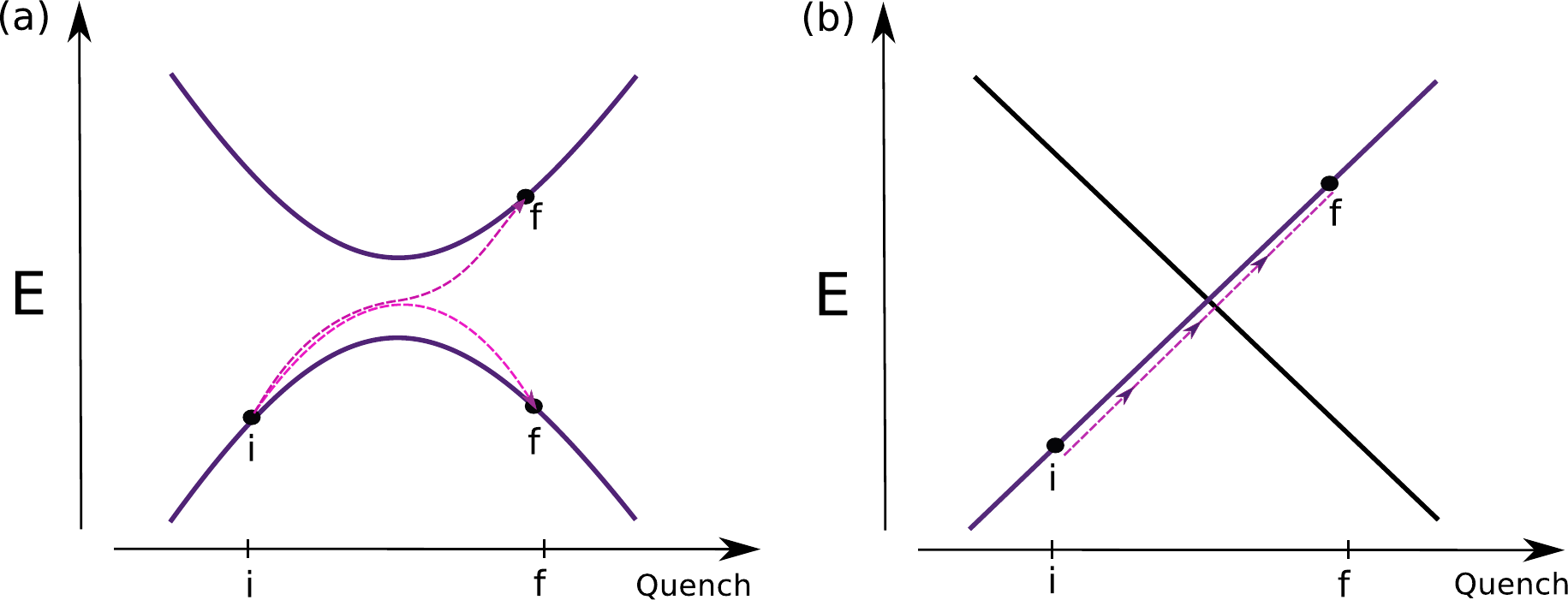}
\caption{(Color online) a) Typical shift of probability amplitude in a Landau-Zener transition for states associated with
generic $(k, -k)$ pairs. b) The level crossing for the occupied $k=0$ state in the odd-fermion
 sector is at the heart of topological blocking; lack of coupling with the unoccupied mode 
and parity constraints force the $k=0$ state to go into the post-quench excited state. } 
\label{fig:MtmEvol} \end{figure}

In the limit $JT \gg 1$, 
Eq.~\eqref{lz} shows that the excitation probability $p_k (\infty)$ is 
equal to 1 for $k = 0$ and $\pi$, and becomes negligible when $k$ deviates 
from those points by an amount which is much larger than $1/\sqrt{\pi JT}$. 
However, for our quench protocol, we see from Eq.~\eqref{tif} that the initial
and final times, $t - a_k$, are functions of $k$; the time $t - a_k = 0$ at 
which the two eigenvalues of the Hamiltonian are separated by the smallest 
amount ($2b_k$) is crossed only if $t_{i,k} < 0$ and $t_{f,k} > 0$,
i.e., if $0 \le k \le \pi/2$. Hence, the excitation 
probability is dominated only by the region near $k = 0$; for exactly
$k=0$, the two-level system undergoes a level crossing and $p_0 = 1$. The modes
near $k=\pi$ never reach the minimum gap region, and for exactly $k=\pi$,
the two-level system remains in the ground state with $p_\pi = 0$.
The behavior of the Landau-Zener transition exhibited by sets of $k$-modes
and the evolution of the special $k=0$ mode is shown in 
Figure~\ref{fig:MtmEvol}. 

In the adiabatic
limit, we see that in the even-fermion sector, if we start in the ground state
given in Eq.~\eqref{gseven} at $h_i = 0$, we reach the ground state in 
Eq.~\eqref{gseven} at $h_f = 2$. However, in the odd-fermion sector, if we 
start in the ground state in Eq.~\eqref{gsodd1} at $h_i = 0$, we 
reach the state in Eq.~\eqref{gsodd1} at $h_f = 0$ which is the ground state
in that sector but which, as discussed above, is separated from the ground 
state of the final Hamiltonian by a finite gap. 
(Note that in the odd-fermion sector, the state with momentum $k = 0$ does 
not change with time since the off-diagonal matrix element $b_0 = 0$ makes it 
impossible to have a transition between the two eigenstates of the 
Hamiltonian). Hence, an 
adiabatic time evolution takes a system from the initial ground state to the 
ground state of the final Hamiltonian in certain sectors but not in others, 
with the different sectors being distinguished from each other by a 
topological quantity, namely, the fermion parity in our model. This 
explicitly demonstrates topological blocking in this system.

{\bf Overlap at the final time:-}
At $t=T$, the overlap between the final state reached and the actual
ground state in a particular sector is given by 
\be {\cal O} (T) = \prod_{0 < k < \pi} (1-p_k (T)). \label{overlap1} \ee
In the limit $JT \gg 1$, we know that $p_k (T)$ is significant only for a 
range of $k$ of the order of $1/\sqrt{\pi JT}$ near $k = 0$. Let us consider 
the thermodynamic limit $N \gg 1$ and define a dimensionless scaling variable 
\be \bar T ~=~ \frac{\pi^2 JT}{N^2}. \ee
Using the fact that the momenta in the even- and odd-fermion sectors are given
by $(2n+1) \pi/N$ and $(2n+2) \pi/N$, where $n= 0,1,\cdots,N/2-1$, we can 
express the overlaps in the even- and odd-fermion sectors as 
\bea {\cal O}_{\text{even}} (T) &\approx & \prod_{n=0}^{\infty} ~\left( 
1 - e^{-\pi (2n+1)^2 \bar T} \right), \non \\
{\cal O}_{\text{odd}} (T) &\approx & \prod_{n=0}^{\infty} ~\left( 1 - e^{-\pi 
(2n+2)^2 \bar T} \right), \label{overlap2} \eea
where we have made the approximation $\sin(k)\approx k$ in Eq.~\eqref{lz}
since only the low lying $k$ modes contribute a significant excitation 
probability. For the same reason we have changed the upper limit from 
$n=N/2-1$ to $\infty$ since
the overlap $1- p_k (T)$ rapidly approaches 1 once $n/N$ becomes a number
of order, say, $0.1$, under the assumption $JT \gg 1$. 

A factor-by-factor comparison of the two expressions in Eqs.~\eqref{overlap2}
shows that ${\cal O}_{\text{odd}}$ 
is larger than ${\cal O}_{\text{even}}$ for any value of $\bar T$. We 
therefore have the interesting result that the overlap between the final 
state and the sectoral ground state is higher in the odd-fermion sector than 
in the even-fermion sector, even though the final state in the odd-fermion 
sector has zero overlap with the ground state of the final Hamiltonian.

We can write the logarithms of the overlaps in Eqs.~\eqref{overlap2} as sums over
$n$. In the limit $\bar T \to 0$, i.e., for $1 \ll JT \ll N^2$, the sums
can be approximated by integrals. Ignoring the difference between $2n+1$ and
$2n+2$ in Eqs.~\eqref{overlap2}, which amounts to ignoring some subleading
terms, we find that in both even- and odd-fermion sectors,
\bea \log {\cal O} (T) &=& \int_0^\infty ~dn ~\log (1 - e^{-4 \pi n^2 \bar T}) 
\non \\
&\approx& - ~\frac{0.653}{\sqrt{\bar T}}. \label{overlap3} \eea

{\bf Overlap at intermediate times:-}
We now look at the overlap between the state reached at a finite time
$t$ and the ground state at that time. This is given by the expression
\be {\cal O} (t) ~=~ \prod_{0 < k < \pi} (1-p_k (t)). \label{overlap4} \ee
As has been analyzed in the context of Landau-Zener
transitions~\cite{Landau32,Majorana32,Vitanov96}, the analytic form of 
$p_k(t)$ can be expressed in terms of Weber functions. 
Numerically we find that for a certain range of values of $\bar T$, the 
overlap ${\cal O} (t)$ of the system shows pronounced oscillations around 
$t=T/2$ (i.e., when $h(t) = 2 Jt/T$ is going through the critical value of 
$J$) before settling down at $t=T$ at a value which is around $0.5$, i.e., 
not very close to either 0 or 1. 
We can estimate this range of values of $\bar T$
by looking at the overlap $1-p_k(t)$ as a function of time $t$ for some 
individual values of the momentum $k$. Assuming that 
$JT \gg 1$, we find the following. For $k \sqrt{JT} \ll 1$ (but not equal to
0), we have an almost sudden process. Hence the overlap stays close to 1 
till we get close to $t=T/2$, and then it rapidly changes to a very small 
value. Clearly, this would make the overlap of the system (which is 
a product of the overlaps for all values of $k$) very small. On the other
hand, for $k \sqrt{JT} \gg 1$, we have an almost adiabatic process and the 
overlap stays close to 1 at all times; such values of $k$ therefore make 
very little difference to the overlap of the system. Only if $k \sqrt{JT}
\approx 0.47$ do we get a final overlap which is around $0.5$. 
(This is consistent with Eq.~\eqref{lz} since $e^{-\pi (0.47)^2} \approx 0.5$).
These different kinds of behavior are shown in Fig.~\ref{fig:overlap} for $JT 
= 100$ and $k \sqrt{JT} = 0.2$, $0.4$, $0.6$ and $0.8$. Thus, the behavior
of the overlap of system that we are looking for, namely, oscillations near 
the critical point before settling down to a value around $0.5$ only occurs
if the {\it smallest non-zero} value of $k$ satisfies $k \sqrt{JT} 
\approx 0.47$. Then this value of $k$ makes the 
dominant contribution to the overlap of the system at all times since all the 
higher values of $k$ contribute factors close to 1 to the overlap. 
Since the smallest non-zero value of $k = m\pi/N$, where $m=1$ and $2$ in
the even- and odd-fermion sectors respectively, the value of $\bar T$ where 
the final overlap of the system is around $0.5$ is about $(0.47)^2 
\approx 0.22$ and $0.22/4 = 0.055$ for even- and odd-fermion sectors 
respectively.

Figure~\ref{fig:overlap} shows oscillations in the overlap near the critical 
region $t=T/2$ which is equal to 50 for our choice of parameters. 
We can understand this by mapping the time evolution with the Hamiltonian in 
Eq.~\eqref{hamlz1} to the Schr\"odinger equation of a particle moving in an
inverted harmonic potential~\cite{Landau32,Vitanov96}. If we define the upper 
and lower components of the two-component wave function associated with
the state $|\psi_k (t,0) \rangle$ by $\psi_{1k}$ 
and $\psi_{2k}$, we can eliminate, say, $\psi_{2k}$ to obtain the equation
\be - ~\frac{d^2 \psi_{1k}}{dt^2} ~-~\left(\frac{4J(t-a_k)}{T}\right)^2 
\psi_{1k} ~ -~ i \frac{4J}{T} \psi_{1k} ~-~ J^2 b_k^2 \psi_{1k} ~=~ 0. 
\label{haminv} \ee
Since we are interested in the behavior of the solution of Eq.~\eqref{haminv}
when $JT k^2 \approx 0.22$ is small (and $a_k \simeq T/2$), we will ignore 
the last term, $J^2 b_k^2 = 4 J^2 \sin^2 (k)$, in comparison with the other 
terms like $4J/T$. The dominant behavior 
of the solutions of Eq.~\eqref{haminv} is then given by $e^{-i2J(t-T/2)^2/T}$.
This explains the oscillations around $t=T/2$. Further, as $t- T/2$ moves away 
from zero, $e^{-i2J(t-T/2)^2/T}$ oscillates more and more rapidly; this is 
qualitatively confirmed by the plots in Fig.~\ref{fig:overlap}.

To summarize the discussion in the last two paragraphs, the overlap
in Eq.~\eqref{overlap4}, in general, either stays close to 1 at all 
times or drops rapidly from 1 to zero when the system crosses the quantum 
critical point at $t=T/2$. The intermediate behavior in which the overlap 
drops to a value which is about halfway between zero and 1 when $t$ 
crosses $T/2$ occurs only when Eq.~\eqref{overlap4} is dominated by the smallest
non-zero value of $k$, and that value of $k$ happens to satisfy
$k\sqrt{JT} \approx 0.47$. For a system of size $N$, the smallest non-zero
value of $k$ is given by $\pi/N$ and $2\pi/N$ in the even- and odd-fermion
sectors, respectively; from this we can deduce the value of $\sqrt{JT}/N$
at which the intermediate behavior occurs in the two sectors. When considered together,
 the highly sensitive nature of this quench behavior on the actual value of momentum, 
the dominance of a single mode in the net overlap, and the slightly different 
momentum quantization conditions for the two sectors, together explain the markedly 
different quantitative behavior shown by the overlap in the two sectors in 
 Figure~\ref{fig:TrIsing_example}. 

\subsection{Other Quantities}
We have found the wave function overlap plotted in 
Fig.~\ref{fig:TrIsing_example} to be the most sensitive yet direct measure 
of the dependence of quench dynamics on topological sectors. In this context, 
we briefly discuss here other quantities that are commonly studied in quench 
dynamics and related Kibble-Zurek physics~\cite{Kibble76,Zurek96,Damski05,
Anatoly05,Levitov06,Mukherjee07,Sengupta08,Mondal08,Degrandi08,Polkov08}. 
In fact, the behavior of several quantities can be traced back to that of 
the probability of excitation within each set of momentum modes, namely, 
that of the $p_k(t)$ which was first introduced in Eq.~\eqref{pkt}.

{\bf Defect density:-} 
The well-studied Kibble-Zurek defect density is the cumulative sum of the 
excitation probabilities for all the modes, i.e. $ n_D \sim \int dk p_k$.
In terms of Ising spins, the defect density is a measure of how many spins 
are pointing in the energetically unfavorable direction in the final phase. 
In the final state reached at $t=\infty$, the total defect density is given by
\be n ~=~ \frac{2}{N} \sum_{k > 0} p_k (\infty). \label{defdens1} \ee
To obtain the standard Kibble-Zurek scaling, in the limit $N \to \infty$, we 
can replace the sum in Eq.~\eqref{defdens1} by an integral and use the 
asymptotic form of $p_k$ given in Eq.~\eqref{lz}, 
\be n ~=~ \int_0^\pi ~\frac{dk}{\pi} ~p_k (\infty)~=~ \int_0^\pi ~
\frac{dk}{\pi} ~e^{-\pi J T \sin^2 (k)}. \label{defdens2} \ee
In the adiabatic limit $JT \to \infty$, only the regions near $k = 0, \pi$ 
contribute to the integral, and we get the Kibble-Zurek scaling law
$ n ~\sim~ T^{-1/2}$. This scaling is exactly mirrored by the behavior of
the logarithm of the overlap ${\cal O}$ in Eq.~\eqref{overlap3}.

As with the overlap, in distinguishing the even and odd sectors, the summation
on $k$ in Eq.~\eqref{defdens1} is restricted to the allowed momenta. The 
defect density is less sensitive than the overlap in distinguishing between 
the different topological sectors for the following reason. If the excitation 
probability $p_k (T)$ is close to 1 for any particular value of $k$, this 
affects the overlap in Eq.~\eqref{overlap1} strongly since it is given by a
product over all $k$ and therefore approaches zero if $1-p_k(T)$ is close to 
zero for any $k$. On the other hand, the defect density in Eq.~\eqref{defdens1}
is given by a sum over all $k$ and is not dominated by any one value of $k$; 
in addition, the sum is divided by $N$ which further reduces the contribution 
from any single value of $k$. 

For a system of finite size $N$, in the topologically blocked odd-fermion 
sector,the special $k=0$ mode has a level crossing and, across the phase transition,
completely evolves into the excited state. Compared to the even sector, this
mode thus contributes a term of order $1/N$ independent of the quench rate.
In the thermodynamic limit, this contribution obviously vanishes while away 
from this limit, the degeneracy in the ferromagnet/topological phase is split 
due to finite size effects. However, in this degenerate phase, the splitting 
is exponentially small as a function of $N$~\cite{Lieb1961}, and is always 
present in numerical simulations and physical systems due to their finite 
size. Thus, observation of the quench-independent $1/N$ jump and its scaling 
behavior of systems size would provide some indication of the difference 
between topological sectors.

{\bf Residual energy:-} 
Another characteristic quantity discussed in quench dynamics is the residual 
energy; this measures the excess energy contained in a post-quench state 
compared to the ground state of the final Hamiltonian. In the transverse 
Ising system, the net residual energy at the end of the quench at time $t=T$
is given by the sum of the contributions of each momentum mode, 
\be {\cal E}_{res, k} ~=~ \langle H_k(T)\rangle ~-~ {\cal E}_{kG} (T), 
\label{defresid} \ee 
where the expectation value of $H_k(T)$ defined in Eq.~\eqref{hamlz1} is
with respect to the time-evolved quench state, and ${\cal E}_{kG} (T)$ is the 
energy of the ground state of $H_k(T)$. 

The arguments made above for the defect density also hold for the residual
energy. It respects the same $T^{-1/2}$ scaling behavior and in considering
the odd- and even-fermion sectors, involves restricted momentum summations. 
As with the defect density, in the odd-fermion sector the $k=0$ makes a 
special contribution, 
taking the time-evolved state completely into the excited branch. Thus, in 
this sector, the residual energy shows a jump of order $J$. This too is an 
effect of order $1/N$ in that there are contributions from a total of $N$ 
momentum sets to the entire residual energy. Nevertheless, the jump reflects 
topological blocking and the difference in behavior of sectors illustrated in
Fig.~\ref{fig:isingchainspec}. 

{\bf Entropies:-} Various forms of entropy, such as the entanglement entropy, 
have been actively studied in the context of quenches. These measures provide an 
alternative picture for the manner in which the wave function evolves. 
In the context of topological sectors, based on the special behavior of the 
$k=0$ mode, i.e., $p_{k=0}(t>0)=1$, we find that a variant of the Renyi 
entropy~\cite{Calabrese10}, $S_{\alpha}$, would provide an 
effective way of distinguishing odd and even sectors:
\be S_{\alpha} ~=~ \frac{1}{1-\alpha} ~\ln ~\large( \sum_{k > 0} ~[ p_k 
(\infty) ]^\alpha \large). \label{eq:entropy} \ee
Given the Kibble-Zurek scaling form discussed above, $S_{\alpha}$ would 
behave as $\ln[\beta_{o/e}+C(\alpha T)^{-1/2}]$, where $C$ is a constant 
and $\beta=0$ for the even-fermion sector while, in the odd-fermion sector, 
$\beta=1$ is derived from the special $k=0$ mode. By picking $\alpha$ to be 
large enough, we could force $C(\alpha T)^{-1/2}\ll 1$, resulting in 
$S_{\alpha}$ being close to zero for the odd-fermion sector and large and 
negative for the even-fermion sector. 

An obviously modified version of this discussion of other quench and 
sector-dependent quantities also holds for the Kitaev model of the 
subsequent section. 

\section{Kitaev's honeycomb model}
\label{sec:Honeycomb}

We now explore a model that is truly topological in that while it possesses 
global topological order and associated degeneracies, it has no local order: 
the Kitaev honeycomb model~\cite{Kitaev2006}, shown in 
Fig.~\ref{fig:orientation2} (see also section~\ref{sec:honeyham} for the full Hamiltonian). The 
model is very rich in and of itself and has the elegant analytic solution 
pioneered by Kitaev as well as various alternate analytic approaches. 
 
Before embarking on the relevant details necessary to analyze the Kitaev model
in the context of our present work, we first outline how our analysis of the 
Kitaev model can be understood as a direct two-dimensional extension of the 
analysis of the previous section. Regardless of whether the reader is 
familiar with the Kitaev honeycomb, this discussion
should make our main results for it clear. 

\subsection{A two-dimensional extension of the transverse Ising chain}
In the previous section, we studied the topological description of the Ising 
chain in terms of a BdG description of a one-dimensional fermionic spinless 
$p$-wave superconductor in a ring geometry. The Hilbert space was divided 
into two sectors consisting of momenta that were quantized either 
according to periodic or anti-periodic boundary conditions and were 
associated with odd- and even-fermion parity, respectively. Depending on 
the parameters in the Hamiltonian, the energetics either allowed
the two sectors to be degenerate in ground state energy or for the odd sector 
to have a higher sectoral ground state energy than that of the even sector. 

With regards to quench dynamics, this mismatch in energy resulted in 
topological blocking in that if one started in the odd sector in the 
degenerate phase and quenched into the non-degenerate phase,
the overlap with the final absolute ground state would be zero. As for 
evaluating overlaps between time-evolved quenched states and the final 
sectoral ground state, this was done by studying the 
simple dynamics of decoupled pairs of momentum states $\pm\bk$. The momenta 
$\bk=0,\pi$ were special since they respect $\bk=-\bk$ and they dictated the
fermion parity. The overlaps clearly showed different behavior that 
depended on the topological (odd/even) sector. 

While the Kitaev honeycomb model has several complex, rich aspects, much 
can be understood by simply generalizing the above to two dimensions. We 
will see that the Kitaev model can be mapped to a spinless two-dimensional 
$p$-wave superconductor and the analog of a ring becomes a torus. Topological 
requirements now dictate periodic or antiperiodic boundary conditions along
the two independent ($x$ and $y$) directions, yielding a total of four 
topological sectors. Unlike in the transverse Ising case, the boundary 
conditions and fermion parity are not simply related. But in the 
commonly-studied situation that the honeycomb system has no vortices, one which 
we confine ourselves to, the fermion parity is constrained to be even. 
As a result, we  find that as a function of parameter space, there 
exist three different phases in which all four sectors have 
degenerate ground states (Abelian $A$ phases). On the other hand, a fourth 
phase (non-Abelian $B$ phase) has its absolute ground state in three of the 
sectors while the fourth sector has higher sectoral ground state energy. 

Thus, similar to the transverse Ising case, topological blocking occurs in 
one out of the four topological sectors. When evaluating overlaps 
between time-evolved quenched states and final sectoral ground states, the 
special momenta are $(k_x,k_y)=(0/\pi,0/\pi)$. In 
Fig.~\ref{fig:overlap_kitaev} we show the typical overlap data for all 4 
sectors over the course of a quench. By symmetry, two of the time-evolved 
overlaps $O_{+-}$ and $O_{-+}$ show identical behaviors. The $O_{--}$ overlap 
is generally different from these other two sectors but this is a finite size 
effect and quickly vanishes for large system sizes. The last overlap $O_{++}$ 
from the fully periodic sector is distinctly higher that the other three. 
This is a consequence of topological blocking. In what follows we will 
explain in more detail the mechanism behind this. 

\subsection{Kitaev honeycomb Hamiltonian}
\label{sec:honeyham}

\begin{figure}
\includegraphics[width=.45\textwidth,height=0.4\textwidth]{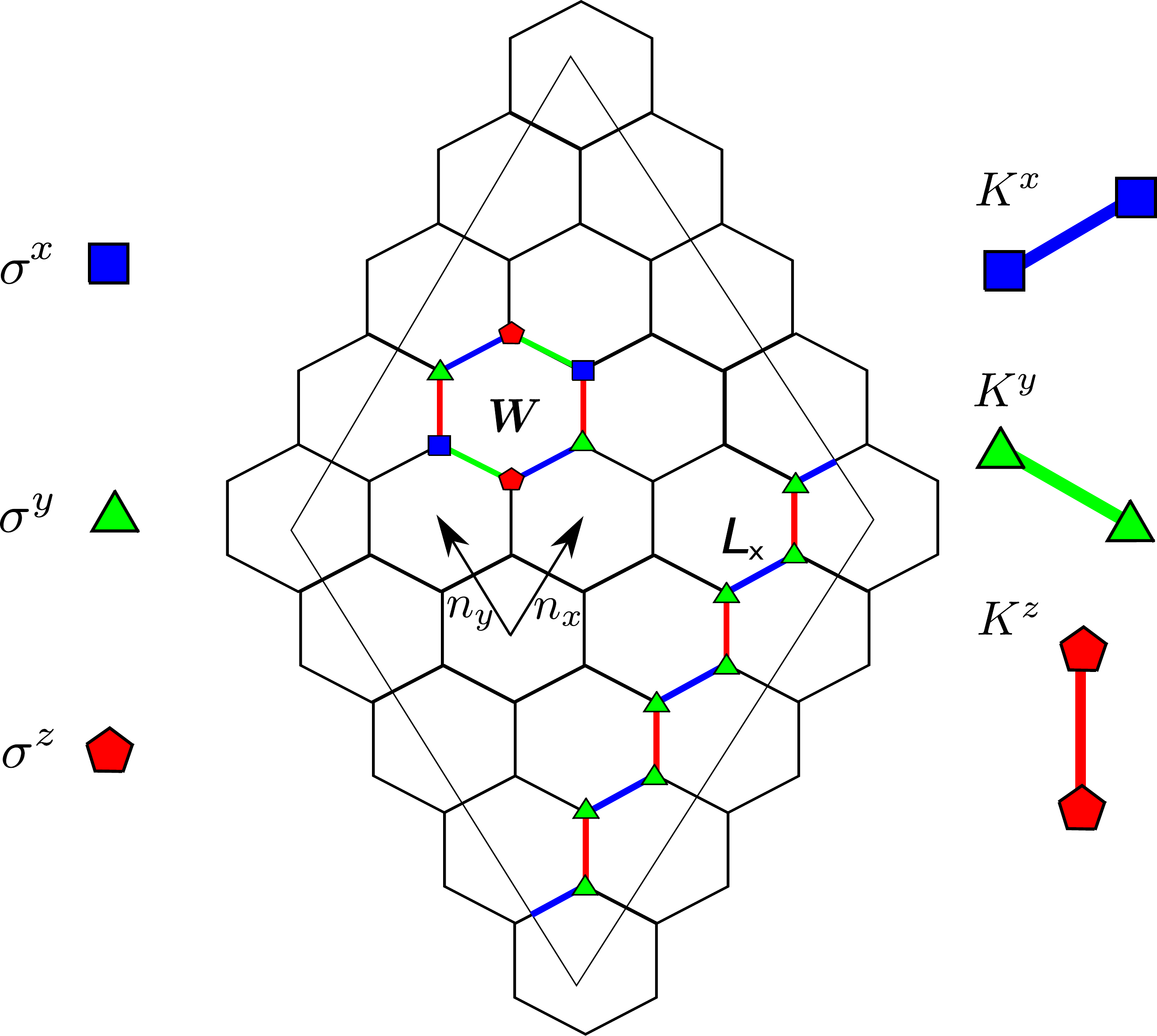}
\caption{(Color online) The honeycomb spin model. The $K^\alpha$ are 
directional spin exchange terms which appear in the Hamiltonian (see section~\ref{sec:honeyham}). On a torus, we identify opposite sides of 
the diamond shape. Symmetries in the model are made by making closed product 
loops of overlapping $K$-terms. The plaquette operators $W$ are the simplest 
symmetries and exist on the surface of the torus. The homologically 
non-trivial symmetries (of which we only indicate $L_x$) are made with 
overlapping products of $K^\alpha$ that loop around the torus.} 
\label{fig:orientation2} \end{figure}

The Kitaev honeycomb system consists of spins on the sites of a hexagonal 
lattice. The Hamiltonian can be written as 
\be H_0 = - \sum_{\alpha \in \{ x,y,z \}} \sum_{i,j} J_\alpha K_{ij}^{\alpha}, 
\label{eq:H} \ee 
where $K_{ij}^\alpha = \sigma_i^\alpha \sigma_j^\alpha$ denotes a directional 
spin exchange interaction occurring between the sites ${i,j}$ connected by a 
$\alpha$-link; see Fig.~\ref{fig:orientation2}. 

Consider now products of $K$ operators along loops on the lattice, 
$K^{\alpha^{(1)}}_{ij} K^{\alpha^{(2)}}_{jk} .... K^{\alpha^{(n)}}_{li}$, 
where $\alpha^{(m)} \in {x,y,z}$. Any loop constructed in this way commutes 
with the Hamiltonian and with all other loops. The shortest such loop 
symmetries are the plaquette operators
\be \bs{W}_{\bq} = \sigma_1^z \sigma_2^x \sigma_3^y \sigma_4^z \sigma_5^x
\sigma_6^y, \label{eq:Wp} \ee
where the numbers $1$ through $6$ label lattice sites on single hexagonal 
plaquette. We will use the convention that $\bq$ denotes the $z$-dimer 
directly below the plaquette. The fact that the Hamiltonian commutes with 
all plaquette operators implies that we may choose energy eigenvectors 
$\ket{n}$ such that $w_{\bq}=\bra{n} \bs{W}_\bq \ket{n}=\pm 1$. If $w_{\bq} 
= -1$ then we say that the state $\ket{n}$ carries a vortex at $\bq$. When 
we refer to a particular vortex-sector we mean the subspace of the system 
with a particular configuration of vortices. The vortex-free sector for 
example is the subspace spanned by all eigenvectors such that $w_{\bq} = 1$ 
for all $\bq$.

On a torus of $N$-spins, there are $N/2$ plaquette ($W_\bq$) operators. 
In general one has the relationship $\prod_{\bq} W_\bq =I$ and so there 
are $N/2-1$ independent plaquette operators. We can find two more 
independent loop operators which we define as overlapping products of 
$K^z$ and $K^x$ or $K^z$ and $K^y$ operators which go around 
homologically non-trivial paths on the torus. We call two such operators, 
which go through the origin, $L_x$ and $L_y$ respectively, see 
Fig.~\ref{fig:orientation2}. We will see that the operators $L_x$ 
and $L_y$ play a role similar to the $T_z$ operator of the one-dimensional
transverse Ising model.

Counting these two operators $L_x$ and $L_y$ together with the plaquettes 
$W_\bq$ gives a total of $N/2+1$ independent symmetries. The different sectors 
are selected by choosing the respective eigenvalues $l_x$, $l_y$ and $w_\bq$. 
The remaining $N/2-1$ degrees of freedom are taken up by $N/2$ fermions (for 
example one for each $K^z$-link) with the constraint on fermionic parity taken into account. 
 
The breaking of $T$-symmetry is essential for relating the model to a chiral 
$p$-wave superconductor. Following the work of Ref.~\onlinecite{Kitaev2006}, 
we use the three-body term 
\be H_1= - \kappa \sum_{\bq} \sum_{l=1}^6 P_\bq^{(l)}, \label{eq:P1} \ee
with the second summation running over the six terms 
\bea \non \sum_{l=1}^6 P_\bq^{(l)} &=& \sigma^x_1 \sigma^y_6 \sigma^z_5 + 
\sigma^z_2 \sigma^y_3 \sigma^x_4 + \sigma^y_1 \sigma^x_2 \sigma^z_3 
\nonumber \\
&& + ~\sigma^y_4 \sigma^x_5 \sigma^z_6 + \sigma^x_3 \sigma^z_4 \sigma^y_5 + 
\sigma^y_2 \sigma^z_1 \sigma^x_6. \label{eq:P2} \eea
For simplicity, in this work we will retain only the terms $P^{(1)},P^{(2)}, 
P^{(3)}$ and $P^{(4)}$.

\subsection{Fermionized solution and phase diagram}

The Kitaev honeycomb Hamiltonian can be solved in several different ways. The 
method implicitly adopted here is the fermionization procedure used in 
Refs.~\onlinecite{Kells2008,Kells2009}. The procedure involves expressing 
the $z$-dimers in terms of hard-core bosons and effective spins and then 
employing string operators to convert bosonic operators to fermionic ones. 
Importantly we can associate the presence of a fermion with an 
antiferromagnetic configuration of the $z-$dimer. 

In the $J_z \gg J_x, J_y$ limit, the ground state manifold contains no 
fermions (spins connected by a $z-$link point in the same direction). The 
remaining degrees of freedom are specified through the eigenvalues of the 
plaquette operators $W$ and the loop operators $L_x$ and $L_y$. It was shown 
by Kitaev [see Ref. \onlinecite{Kitaev2006}] that this manifold can be 
perturbatively mapped on the $4^{\text{th}}$ order to a toric code Hamiltonian
\be H_{TC} = E_0 -\frac{J_x^2 J_y^2}{16 |J_z|^3} \sum Q_\bq, 
\label{eq:TC}\ee
with $Q_\bq = P [W_\bq]$ where $P$ is the projector to the ferromagnetic 
subspace. In this limit, because the projector preserves the eigenvalues of 
$W$ and $Q$ and because the operators $L_x$ and $L_y$ do not appear, there 
are four ground states (labeled by the eigenvalues $l_x$ and $l_y$) with no 
vortices. As the relative values of $J_x$ and $J_y$ become larger, the ground 
states acquire non-zero fermionic components. However, the overall parity 
of these states cannot change and it can be proved that the ground states are 
always vortex-free~\cite{Lieb1994}. Hence, given that the zero vortex sector 
in the toric code limit has no fermions, in the full Kitaev model, this sector, 
which contains the ground state, has even parity. 

In the vortex-free sector of the Kitaev model, $w_\bq=1$ $\forall \bq$, 
and the associated translationally invariant Hamiltonian can be expressed in 
momentum space. In terms of fermionic momentum-space operators $c_{\bk}$,
 the Hamiltonian takes the BdG form~\cite{Kells2009}
\be H= \frac{1}{2} \sum_{\bk} \left[\begin{array}{cc} 
c^\dagger_{\bk} & c_{-\bk} \end{array} \right] H_\bk \left[\begin{array}{c}
c_{\bk} \\ c^\dagger_{-\bk} \end{array} \right], \label{eq:Hm1} \ee
with
\be H_\bk=\left[ \begin{array}{cc} 
\xi_\bk & \Delta_\bk \\ \Delta_\bk^* & -\xi_\bk \end{array} \right],
\label{eq:Hk1} \ee
where 
\bea \xi_\bk &=& \varepsilon_\bk -\mu, \non \\
\Delta_\bk &=& \alpha_\bk+i \beta_\bk, \eea
and
\bea \mu_{\phantom \bk} &=& -2 J_z, \non \\
\varepsilon_\bk &=& 2J_x \cos(k_x) + 2J_y \cos(k_y), \non \\
\alpha_\bk &=& 4 \kappa [ \sin(k_x) -\sin(k_y)], \non \\ 
\beta_\bk &=& 2 J_x \sin(k_x) + 2 J_y\sin(k_y). \label{parameters} \eea
Here, $\bk$ denotes the two-dimensional vector given by momentum 
components $(k_x,k_y)$.
Thus, the Kitaev honeycomb system maps to a spinless fermionic BdG 
Hamiltonian, which when compared to that associated with the transverse Ising
chain in the previous system, can be regarded as a two-dimensional 
extension. All terms in $\Delta_\bk$ carry net angular momentum $l=1$ and thus
the superconducting gap is of $p$-wave nature. The three-body terms in 
Eqs.~(\ref{eq:P1}-\ref{eq:P2}) can be seen to open the gap in the $B$ phase of the model and provide a $T$-symmetry breaking component that makes the system chiral. 

As in Sec.~\ref{sec:IsingFermionic} for the 1D case, we diagonalize 
the BdG Hamiltonians $H_\bk$ by defining the Bogoliubov-Valatin operators
\bea \label{eq:gamma2}
\gamma_{\bk} &=& u_\bk c^{\phantom \dagger}_\bk - v_\bk c^{\dagger}_{-\bk}, 
\non \\
\gamma_{\bk}^\dagger &=& u_\bk^* c^\dagger_\bk - v_\bk^* c^{\phantom 
\dagger}_{-\bk}, \eea
with
\bea \label{eq:uv2}
u_\bk &=& \phantom -\sqrt{(1 + \xi_\bk/\epsilon_\bk)/2}, \non \\
v_\bk &=& -\sqrt{(1 - \xi_\bk/ \epsilon_\bk)/2} ~~e^{i \arg (\Delta_\bk)}. \eea
As with the 1D case, the modes with $k_x, k_y = 0$ and $\pi$ require a special
analysis since they satisfy $k = - k$. Further, $\Delta_k = 0$ for these 
modes; hence, $\varepsilon_k = |\xi_k|$. The diagonalized Hamiltonian once more
 takes the form
\bea H &=& \sum_\bk \epsilon_\bk (\gamma^{\dagger}_{\bk} 
\gamma^{\phantom \dagger}_\bk -1/2), \label{eq:Hdiag2} \non \\
\epsilon_\bk &=& \sqrt{\xi_\bk^2 +|\Delta_\bk|^2}. \label{xidelta2} \eea
The ground state of this has the BCS form,
\be \ket{\text{gs}} = \prod_\bk (u_\bk + v_\bk \crc_\bk \crc_{-\bk} )
\ket{\text{vac}}, \label{eq:BCS1} \ee
which is annihilated by all the $\gamma_\bk$, and has the energy 
$E_{\text{gs}}=-\frac{1}{2}\int E_\bk d \bk $.

The form of the dispersion in Eq.~\eqref{xidelta2} enables us to derive the 
phase boundaries and gapped/gapless nature of the phases in the honeycomb 
system. We assume that $J_x, J_y, J_z > 0$. As we mentioned above, with this 
convention the $c$ fermions are associated with antiferromagnetic 
configurations of the $z$-dimers and our vacua are toric code states on an 
effective square lattice~\cite{Kells2009}. 

We first consider the case $\kappa=0$; then Eqs.~\eqref{parameters} 
are the same as those used in previous work involving quenches in the 
Kitaev honeycomb model, namely Ref.~\onlinecite{Sengupta08} with 
$\vec{M}_1=k_x$ and $\vec{M_2}=-k_y$. 
{}From the dispersion, it can be seen that the system is gapless in the range
$|J_x-J_y|<J_z<J_x+J_y$, and by symmetry, within similar constraints on
$J_x$ and $J_y$. Thus, as was originally discussed by Kitaev, the system 
has four phases~\cite{Kitaev2006}. The system is gapped in three of the 
phases, $A_x$, $A_y$ and $A_z$, having $J_x > J_y + 
J_z$, $J_y > J_z + J_x$, and $J_z > J_x + J_y$ respectively. 
These are called Abelian phases because the
low-energy excitations satisfy Abelian statistics under exchanges.
In the fourth phase, called $B$, each of the $J_i$ is less than the
sum of the other two couplings. The spectrum is gapless in this phase.
(This makes it difficult to compute the statistics of the low-energy 
excitations since even a slow exchange of two of them inevitably produces 
other low-energy excitations). For instance, if $J_x = J_y = J > J_z/2$
and $\kappa = 0$, we find that the spectrum is gapless at two points given by
$k_x = - k_y = \pm \cos^{-1} (-J_z/2J)$. The spectrum close to these points 
has the gapless Dirac form with the Dirac cones touching at those points. 

If we now make $\kappa \ne 0$, phase $B$ also becomes gapped, with the 
minimum gap occurring at the two points mentioned above if $\kappa$ is small.
The low-energy excitations in this phase are then found to satisfy non-Abelian
statistics. 

\begin{figure}
\includegraphics[width=0.45\textwidth,height=0.35\textwidth]{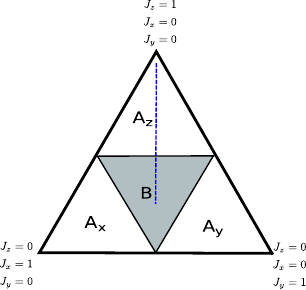}
\hspace*{.02cm} 
\caption{Schematic of the Kitaev model phase diagram. The $A$ regions 
correspond to toric code phases while the $B$ region corresponds to the 
non-Abelian Ising phase. The dashed line indicates the quench parametrization. 
Note that this diagram uses Kitaev's normalization $J_x+J_y+J_z=1$. In our quench protocol, we use $J_x=J_y=1$ with $J_z$ running between $1$ and $3$ but on renormalizing this gives a path similar to the dashed line shown in the diagram.
 }
\label{fig:phasemap} \end{figure}

The four phases are separated by quantum phase transition lines on which one 
of the $J_i$ is equal to the sum of the other two couplings. As is standard, 
the four phases can be depicted in the triangular phase diagram shown in 
Fig.~\ref{fig:phasemap}. 

\subsection{Topological degeneracy}

The topological nature of the four phases can be directly gleaned by putting 
the system on a torus. We discussed briefly above how the four-fold degeneracy 
of the $A$-phases could be understood by mapping perturbatively to the 
toric code. Let us now see how this looks within the exact fermionic 
solution of the model where we can also understand the three-fold degeneracy 
of the non-Abelian phase and the gapless nature of the blocked sector.

We remark here that while most of the analysis for the transverse Ising 
system can be extended into two dimensions for the Kitaev honeycomb system, 
one crucial difference occurs with regard to fermion parity. In the Ising system,
two sectors were allowed based on fermion parity, namely odd and even sectors,
 and while these were degenerate in one phase, they were not so in the other. 
Here, all states in the vortex-free sector have even fermion number parity, as argued after Eq.~\eqref{eq:TC}. 
As we shall see below, the degeneracies come about from different 
combinations of even fermion occupation.
 
We assume that the number of sites in the $x$ and $y$ directions are $N_x$ and 
$N_y$, with the first site linked to the $N_i$th site along each direction. 
On the torus, the diagonalized Hamiltonian has a restricted set of 
momentum modes in its form
\be H_ = \sum_{k_x,k_y} E_{\bk}( \gamma^\dagger_\bk
\gamma_\bk^{\phantom \dagger} - \frac{1}{2}), \ee
where the dispersion relation $E_\bk$ is given in Eq.~\eqref{xidelta2}. 
The allowed values of $k_\alpha$ in the various homology sectors on the torus 
are $\theta_\alpha + 2 \pi \frac{n_\alpha}{N_\alpha}$ for integer $n_\alpha
=0,1,...N_\alpha-1$, where the four topological sectors corresponding to 
$(l_x,l_y)=(\pm 1,\pm 1)$ have values of $\theta_{\alpha}$ given 
by $\theta_\alpha = (\frac{l_{\alpha}+1}{2})\frac{\pi}{N_\alpha} $. The 
topological sectors dictate whether the wave functions are periodic or 
antiperiodic. The relationship between the topological sectors and the 
periodicity/antiperiodicity of the wave functions is simple if a 
little counter intuitive. For example the fully periodic sector $(+,+)$ 
has the quantum numbers $(l_x,l_y)=(- 1,- 1)$, while the fully 
antiperiodic sector $(-,-)$ has quantum numbers $(l_x,l_y)=(1,1)$.

We know that the ground state in the vortex-free sector has even-fermion 
parity. It can then be shown that in the three topological sectors 
corresponding to $(l_x,l_y) = (1,1)$, $(-1,1)$ and $(1,-1)$, the 
momenta $\bk = (0,0)$, $(0,\pi)$, $(\pi,0)$ and $(\pi,\pi)$ are missing, and
the ground state is always of the form given in Eq.~\eqref{eq:BCS1} with the 
momenta discretized appropriately as described above. As parity is conserved 
in each vortex sector, the eigenstates above these ground states are reached 
by creating fermions in pairs.

In the fully periodic sector $(+,+)$ things are not as straightforward because
the four momenta $\bk = (0,0)$, $(0,\pi)$, $(\pi,0)$ and $(\pi,\pi)$ are 
present. Since $\bk = - \bk$ and $\Delta_\bk = 0$ for all these modes, we have
the energetics for these states being purely determined by $\varepsilon_\bk 
- \mu = 2(J_x\cos(k_x)+J_y\cos(k_y)+J_z)$. In particular, we
have $u_\bk = 1$, $v_\bk = 0$ and $\gamma_\bk = c_\bk$ if $\varepsilon_\bk - 
\mu > 0$, while $u_\bk = 0$, $|v_\bk| = 1$, and $\gamma_\bk = c_\bk^\dagger$ 
if $\varepsilon_\bk - \mu < 0$; in either case $E_\bk = |\varepsilon_\bk - 
\mu| > 0$. 

The situation in the four different phases and for the four 
different momenta is summarized in Table~\ref{table:phases_momenta}. In the 
Abelian phase $A_z$, where $J_z>J_x+J_y$, we have $\gamma_\bk = c_\bk$ for 
all the four momenta and we can use the BCS ground state in 
Eq.~\eqref{eq:BCS1} where all these momenta are excluded. In the phase 
$A_x$, where $J_x>J_y+J_z$, we have $\gamma_\bk \sim 
c_\bk^\dagger$ for $\bk = (\pi,0)$ and $(\pi,\pi)$, so these two momenta must 
be included as a factor $c_{\pi,0}^\dagger c_{\pi,\pi}^\dagger$ in 
Eq.~\eqref{eq:BCS1}. Similarly, in $A_y$, where $J_y>J_z+J_x$, we have 
$\gamma_\bk \sim c_\bk^\dagger$ for 
$\bk = (0,\pi)$ and $(\pi,\pi)$, so these momenta must be included as a factor
$c_{0,\pi}^\dagger c_{\pi,\pi}^\dagger$ in Eq.~\eqref{eq:BCS1}. We see that 
in the three Abelian phases, an even number of momenta are included so that 
the fermion parity is even as required. We remark here that the asymmetry in 
the momentum occupation structure between $A_z$, and $A_x$ and $A_y$ can be 
traced back to the original transformations of the honeycomb Hamiltonian 
involving dimerization in the $z$-bonds. While this structure is basis 
dependent, topological aspects, such as degeneracies, are not. 

In the $B$ phase, we have $\gamma_{\pi,\pi} = c_{\pi,\pi}^\dagger$. However, 
we cannot include a factor of $c_{\pi,\pi}^\dagger$ in Eq.~\eqref{eq:BCS1} by itself,
since this would make the fermion parity odd.
In Ref.~\onlinecite{Kells2009} it was demonstrated that the states 
\be \label{eq:Ezero}
\ket{\psi}_0= \prod_{\bk \neq (\pi,\pi)} (u_\bk +v_\bk c^\dagger_\bk 
c^\dagger_{-\bk} ) \ket{\text{vac}} \ee
and
\be 
\label{eq:Bband}
\ket{\psi}_{\bk'} =c_{\pi,\pi}^\dagger \gamma^\dagger_{\bk'} 
\prod_{\bk \neq (\pi,\pi)} (u_\bk +v_\bk c^\dagger_\bk
c^\dagger_{-\bk} ) \ket{\text{vac}} \ee
can be used as replacements in this scenario, where $\bk'$ can be any
momentum apart from $(\pi,\pi)$, thus forming a band. Since the dispersion 
relation is gapped, all of these states have a higher energy than the ground 
states of the other three topological sectors. 

\begin{table}[ht] \centering \begin{tabular}{|c|c|c|}
\hline
Phase & $\varepsilon_\bk - \mu > 0$ & $\varepsilon_\bk - \mu < 0$ \\
\hline
$A_x$ & $(0,0)$, $(0,\pi)$ & $(\pi,0)$, $(\pi,\pi)$ \\
$A_y$ & $(0,0)$, $(\pi,0)$ & $(0,\pi)$, $(\pi,\pi)$ \\
$A_z$ & $(0,0)$, $(\pi,0)$, $(0,\pi)$, $(\pi,\pi)$ & --- \\
$B$ & $(0,0)$, $(\pi,0)$, $(0,\pi)$ & $(\pi,\pi)$ \\
\hline
\end{tabular}
\caption{The special momenta $\bk$ for which $\Delta_\bk=0$ and either 
$\varepsilon_k - \mu > 0$, $\gamma_\bk = c_\bk$ or $\varepsilon_k - \mu < 0$, 
$\gamma_\bk = c_\bk^\dagger$, in the four phases $A_x$, $A_y$, $A_z$ and $B$. 
This structure belonging to the fully periodic topological sector determines 
the ground state degeneracies of each of the phases.}
\label{table:phases_momenta} \end{table}

We can summarize the situation in the thermodynamic limit $N_x, N_y \gg 1$ 
as follows. In the three Abelian phases, the ground states in all the four 
topological sectors are degenerate with each other; hence the ground state of 
the system has a four-fold degeneracy. However, in the $B$ phase, the ground 
state in the three topological sectors with $(l_x,l_y) = (1,1)$, $(-1,1)$ 
and $(1,-1)$ are degenerate with each other, while the ground state in the 
sector $(l_x,l_y) = (-1,-1)$ has a higher energy; hence the ground 
state of the system has a three-fold degeneracy. The situation is similar to
the transverse Ising model where the ground state has a two-fold degeneracy
in the ferromagnetic phase and has no degeneracy in the paramagnetic phase. 
We therefore expect that the Kitaev model on a torus will also exhibit 
topological blocking when the parameters in the Hamiltonian are quenched so 
as to take it from any one of the Abelian phases to the $B$ phase. 

\subsection{Quenching Dynamics}

We will now study the quenching dynamics on a torus and 
discuss the ground state overlap $O_{\pm,\pm}(t)$ for each of the four 
topological sectors as we quench from one of the Abelian phases through a 
phase transition line into the $B$ phase. As in the previous section, we first 
discuss topological blocking, then the qualitative features for the time 
evolution of momentum modes, and then detailed numerical results.

\begin{figure}
\includegraphics[width=0.45\textwidth,height=0.25\textwidth]{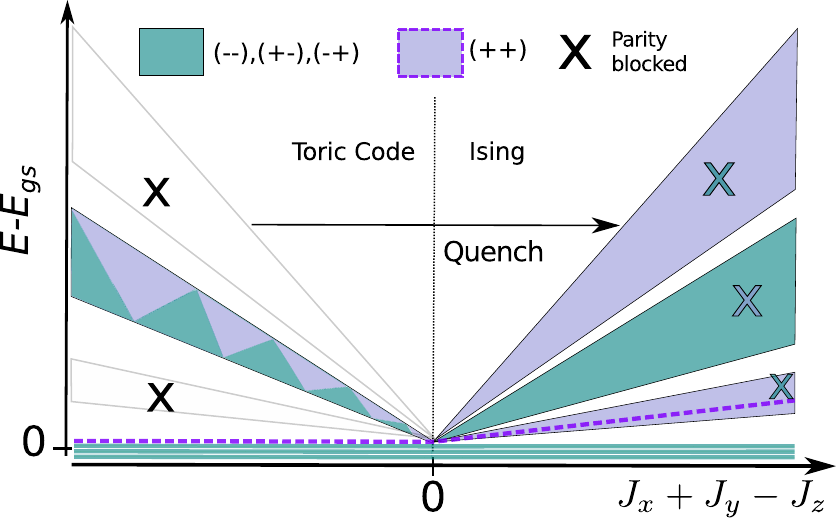}
\hspace*{.02cm} 
\caption{ (Color online) Schematic of the spectrum of the vortex free sector of the honeycomb 
model as a function of $J_x+J_y-J_z$. 
In the toric code phase ($J_z> J_x+J_y$), the ground state is four-fold 
degenerate in the thermodynamic limit with all states being constructed 
from an even number of fermion excitations. The excitation spectrum consists 
of bands of states created from the four ground states using pairs of 
$\gamma^{\dagger}$ operators. As in the transverse Ising case, there are no 
energy levels with an odd number of $\gamma^{\dagger}$ excitations over one 
of the ground states. In the non-Abelian Ising phase ($J_z< J_x+J_y$), there 
is a three-fold degenerate ground state. The parity blocking mechanism means 
that the fully periodic sector is gapped away from the other three states and that
the lowest lying states in this sector form part of a band.  Further bands are created from the ground states and this lowest band using pairs of $\gamma^{\dagger}$ operators. The purple dashed line indicates that, in the adiabatic limit, the (++) sectoral ground state in the Toric Code phase flows into this lowest band, but does {\em not} necessarily flow to the lowest energy state in the (++) sector of the Ising phase (the lowest state in the band usually occurs at a different momentum from the initial state and our quench conserves momentum)} \label{fig:topblockhc} \end{figure}

Topological blocking in this case is in principle also straightforward , see Figure \ref{fig:topblockhc}. The ground state with $(l_x,l_y)=(-1,-1)$, which exists in the $A$ phases, does not have a counterpart ground state in the $B$ phase, but the topological quantum numbers $l_x$ and $l_y$ are conserved in a quench. Therefore, when quenching from the $(l_x,l_y)=(-1,-1)$ ground state of an $A$ phase into the $B$ phase, the system is blocked from reaching any of the ground states of the $B$ phase. Understanding this blocking in terms of the occupation of momentum modes is more subtle in the Kitaev honeycomb system than in
the transverse Ising system, as can be surmised from the discussion of
degeneracies in the previous subsection. In the three sectors having 
$(l_x,l_y) = (1,1)$, $(-1,1)$ and $(1,-1)$, nothing strange 
occurs since the four special momenta do not exist in those sectors. 
But in the sector $(+,+)$ where $(l_x,l_y) = (-1,-1)$,
the four momenta exist and they do not evolve at all with time as they 
do not mix with any other momenta (since $\Delta_\bk = 0$ for these modes).
We can now understand what will happen to the ground states of the Abelian phases as we quench into the $B$ phase by looking 
at Table~\ref{table:phases_momenta}. If we start in 
the ground state of $A_x$, which has the modes with momenta
$(\pi,0)$ and $(\pi,\pi)$ occupied, and we
quench across the line $J_x = J_y + J_z$, we will reach
a low-lying state of the $B$ phase which still has these modes occupied. The lowest energy state we can reach is the state in Eq.~\eqref{eq:Bband} with $\bk'=(\pi,0)$.  If we start in the ground state in $A_y$, which has the modes with 
momenta $(0,\pi)$ and $(\pi,\pi)$ occupied, and we quench across the line $J_y = J_z + J_x$,
 we will again reach
a low-lying state of the $B$ phase with these modes still occupied. The lowest energy state we can reach is now the state in Eq.~\eqref{eq:Bband} with $\bk'=(0,\pi)$.
Finally, if we start in the ground state of phase 
$A_z$, the mode with momentum $(\pi,\pi)$ is unoccupied. If we then quench into phase $B$, the lowest state of the $B$ sector that we can reach is the state in  Eq.~\eqref{eq:Ezero}.  Of course none of the states we reach in this way are ground states of the $B$ phase - the actual ground states are in the other $(l_x,l_y)$ sectors. In fact, the lowest states we can reach are not even the lowest energy states of the low lying band in the $(+,+)$ sector in the  $B$ phase. This is because during the quench, momentum is conserved (the Hamiltonian is always translationally invariant in space) and the lowest states in this band occur at different momenta than the ground states of the $A$ phases in the $(+,+)$ sector. Of course, by changing the details of the quench, breaking translational invariance, we should be able to arrange that the system will flow into the lowest states of the band, but unless we introduce non-local perturbations, we will not be able to change the quantum numbers $l_x$ and $l_y$. 

In light of the above discussions, we now study quench dynamics within 
different topological sectors by initializing the system to a ground state in
one of the $A$ phases and quenching through a critical point into the $B$ 
phase. Our specific quench protocol respects the following evolution:
\bea J_x = J_y = J, && \kappa = 0.1J, \nonumber \\
J_z(t) = J(3 - 2t/T), && {\rm for} ~~0<t<T. \label{eq:quenchKitaev} \eea
Thus, we start at $t=0$ at $J_z=3J$, which lies in the $A_z$ phase,
and we end at $t=T$ at $J_z = J$ which lies exactly in the middle of the $B$ 
phase. The phase transition occurs at $t=T/2$ when $J_z = J_x + J_y$.
We note that the since the Hamiltonian conserves the quantum numbers 
$(l_x,l_y)$, the calculations in the different sectors are 
independent of each other and there is no mixing between topological sectors.

The analysis of the quenching problem here is very similar in principle to that in the
transverse Ising model. Each pair of values $\bk$ and $-\bk$ forms a coupled
two-level system having Landau-Zener type dynamics, save for the four special
 momenta $(0,0)$, $(0,\pi)$, $(\pi,0)$ and $(\pi,\pi)$ which require a 
special analysis as we have discussed above. The dynamic Hamiltonian 
appropriate for each two-level set takes the form
\be H (t) ~=~ J~ \left( \begin{array}{cc}
-4(t-a_\bk)/T & b_\bk \\
b_\bk^* & 4(t-a_\bk)/T \end{array} \right), \label{hamlz2} \ee
where 
\bea a_\bk &=& \frac{T}{2} ~(3 + \cos (k_x) + \cos (k_y)), \non \\
b_\bk &=& (2i + 4 \kappa) ~\sin (k_x) ~+~ (2i - 4 \kappa) ~\sin (k_y). 
\label{tauab2} \eea
During the quench, the gap goes through zero only near 
$\bk = (\pi,\pi)$. Hence, if $JT \gg 1$, the transition probability 
$p_\bk$ will differ substantially from zero only in that momentum region.
Since the initial and final times $-a_\bk$ and $T-a_\bk$ are approximately
given by $-T/2$ and $T/2$, and $T$ is large, we can use the expression in 
Eq.~\eqref{lz} for $p_\bk$.
Assuming that $N_x = N_y = N \gg 1$, we define the variable
$\bar T = \pi^2 JT/N^2$. Expanding the momentum around $\bk = (\pi,\pi)$,
we have the following expressions in the four topological sectors,
\bea && (\pi - k_x, \pi - k_y) \non \\
&& =~ \begin{cases} \frac{\pi}{N} ~(2n_x+2,2n_y+2) ~~~{\rm in}~~~(+,+), \\
\frac{\pi}{N} ~(2n_x+2,2n_y+1) ~~~{\rm in}~~~(+,-), \\
\frac{\pi}{N} ~(2n_x+1,2n_y+2) ~~~{\rm in}~~~(-,+), \\
\frac{\pi}{N} ~(2n_x+1,2n_y+1) ~~~{\rm in}~~~(-,-), \end{cases} \non \\
&& \label{nxny} \eea
where $n_x = 0,1,\cdots,N/2-1$ and $n_y = -N/2, -N/2 +1, \cdots, N/2-1$.
We have chosen these ranges of $n_x$ and $n_y$ in such a way that for each
pair of values $\bk$ and $-\bk$, exactly one value of $\bk$ appears in
Eq.~\eqref{nxny}. 

At this point we recall the subtlety of topological blocking in the 
fully periodic sector $(+,+)$ of the Kitaev model. Namely, if we start in any 
of the $A$ phases, a quench through the critical point will not take us to 
the ground state of the $B$ phase because one of the four special momenta
($(0,0)$, $(\pi,0)$, $(0,\pi)$ and $(\pi,\pi)$) will fail to change to its
appropriate ground or excited state. 

Using Eqs.~\eqref{nxny} to write $\bk$ in terms of $n_x,n_y$,
we find that the overlaps between the ground state and the state reached 
at the final time $t=T$ in the different sectors are given by 
\bea && {\cal O}_{\pm,\pm} (T) ~=~ \prod_{n_x=0}^\infty 
\prod_{n_y=-\infty}^\infty ~(1 ~-~ p_{n_x,n_y}), \non \\
&& {\rm where} ~~~ p_{n_x,n_y} ~=~ e^{-\pi JT |b_{n_x,n_y}|^2 /4} \non \\
&& =~ \begin{cases} 
e^{- \pi {\bar T} [(2n_x + 2n_y + 4)^2 + 4 \kappa^2 (2n_x - 2n_y)^2]} ~~~
{\rm in}~~~(+,+), \\
e^{- \pi {\bar T} [(2n_x + 2n_y + 3)^2 + 4 \kappa^2 (2n_x - 2n_y +1)^2 
]} ~~~{\rm in}~~~(+,-), \\
e^{- \pi {\bar T} [(2n_x + 2n_y + 3)^2 + 4 \kappa^2 (2n_x - 2n_y -1)^2 
]} ~~~{\rm in}~~~(-,+), \\
e^{- \pi {\bar T} [(2n_x + 2n_y + 2)^2 + 4 \kappa^2 (2n_x - 2n_y)^2 
]} ~~~{\rm in}~~~(-,-). \end{cases} \non \\
&& \label{overlap5} \eea
Note that we have changed the upper limit for $n_x$ from $N/2-1$ to $\infty$
and the limits for $n_y$ from $[-N/2-1,N/2]$ to $[-\infty,\infty]$; this is
justified for large values of $T$ since the overlap $1- p_{n_x,n_y} (T)$
rapidly approaches 1 once $n_x/N, |n_y|/N$ become numbers of order 1. 
For $4 \kappa^2 < 1$, a term-by-term comparison shows that ${\cal O}_{+,+} >
{\cal O}_{+,-} = {\cal O}_{-,+} > {\cal O}_{-,-}$.

As in the transverse Ising model, the log of the overlaps can be written as
sums over $n_x, n_y$, which can then be written as integrals in the limit
$\bar T \to 0$. Ignoring the integers $1,2,3,4$ in Eqs.~\eqref{overlap5} 
which amounts to ignoring some subleading terms, we find that 
\bea && \log {\cal O} (T) \non \\
&& = \int_0^\infty \int_{-\infty}^\infty dn_x dn_y ~\log (1 - e^{-4 \pi 
{\bar T} [(n_x + n_y)^2 + 4 \kappa^2 (n_x - n_y)^2 ]}) \non \\
&& = - ~\frac{\pi^2}{192\kappa \bar T} \label{overlap6} \eea
in all four topological sectors. We observe that this diverges if $\kappa
\to 0$; this is because in this limit, the off-diagonal element in the
Hamiltonian in Eq.~\eqref{hamlz2} vanishes at not just the four special
momenta but along the entire line in the Brillouin zone given by $k_y = - 
k_x$. The space of momenta for which $p_\bk = 1$ is therefore no longer zero 
dimensional, but instead one dimensional. Looking at the expressions
in Eqs.~(\ref{overlap5}-\ref{overlap6}), we see that if $\kappa = 0$, 
the integral over $n_x + n_y$ gives a factor of order $1/\sqrt{\bar T}$ 
while the integral over $n_x - n_y$ gives a factor of order $N$. We 
therefore expect that $\log {\cal O} (T)$ will be of order $N^2 /\sqrt{JT}$
if $\kappa = 0$; this is to be contrasted with the term of order $1/(\kappa 
{\bar T}) \sim N^2/(\kappa JT)$ that we get if $\kappa \ne 0$. We note here 
that the Kibble-Zurek power-law for the defect density is 
known~\cite{Sengupta08} to have a similar dependence on $\kappa$; the power-law
changes from $T^{-1/2}$ for $\kappa = 0$ to $T^{-1}$ for $\kappa \ne 0$.

\begin{figure}
\includegraphics[width=0.45\textwidth,height=0.35\textwidth]{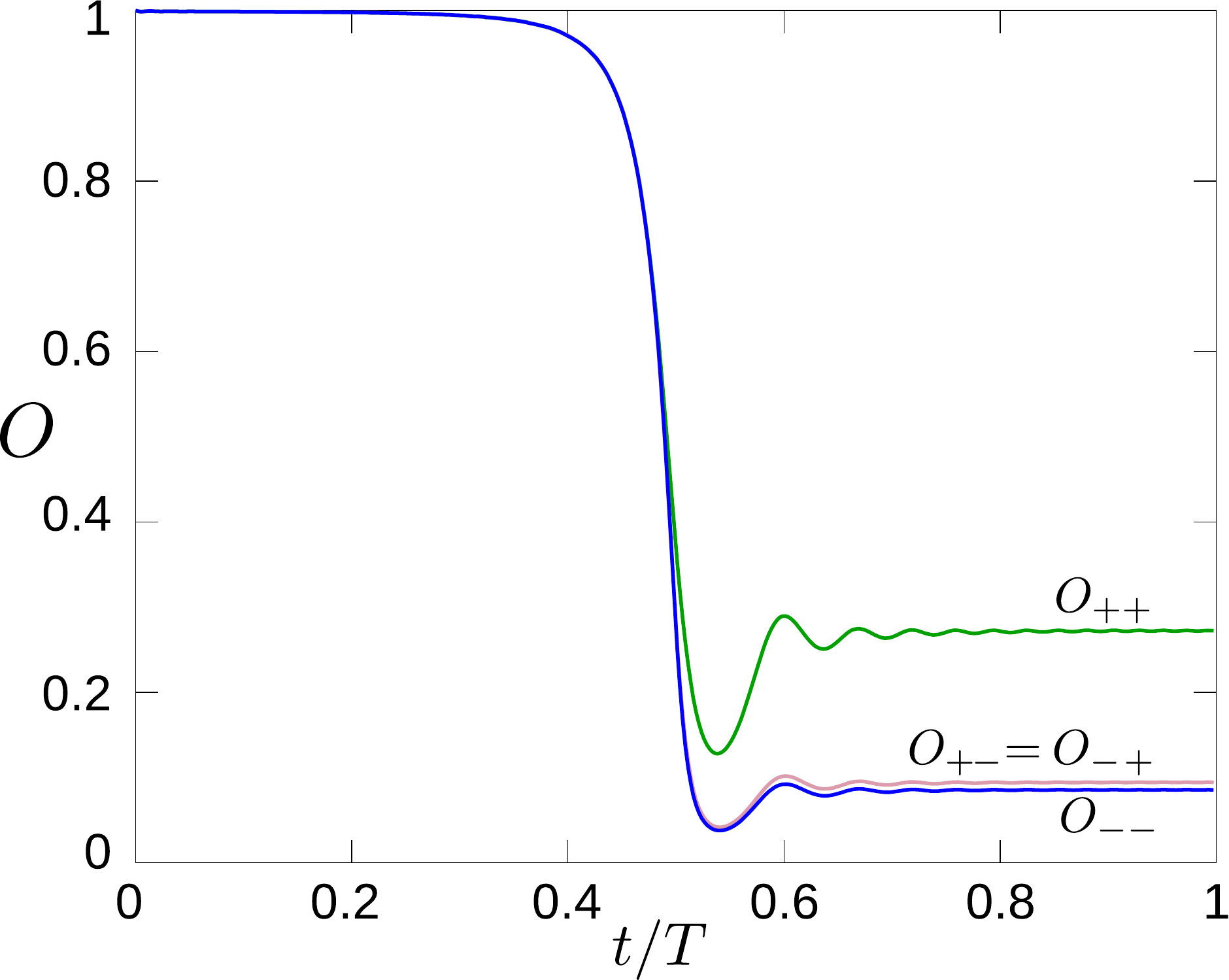}
\hspace*{.02cm} 
\caption{Typical overlap quench data for Kitaev spin model. In this example 
we examined a torus of $N_x=N_y=100$ with a fixed $J_x=J_y=1$, $\kappa=0.25$ 
and time dependent $J_z = 3 - (2 t/T) $ with $T=80$. The blocked sector (++) 
has a generically higher end overlap. Small differences between the other 
sectors are due to finite system size.} \label{fig:overlap_kitaev} \end{figure}

As in the transverse Ising model, the overlaps ${\cal O}_{\pm,\pm} (t)$
show oscillations around $t= T/2$ for a certain range of values of $\bar T$. 
As before, this can be understood by looking at the overlaps for individual 
values of $\bk$ lying in the region 
close to $(\pi,\pi)$. For $\bar T \ll 1$, the overlap changes quickly from 1 
to small values for several values of $\bk$ in that region; hence the 
overlap of the system (which is given by the product of the overlaps over all 
values of $\bk$) changes rapidly from 1 to a very small value. For $\bar T 
\gg 1$, the overlap remains close to 1 for all values of $\bk$; hence the 
overlap of the system remains close to 1. Thus the overlap shows noticeable
oscillations near $t=T/2$ only if $\bar T$ has a value of order 1 in such a 
way that only the value of $\bk$ lying closest to $(\pi,\pi)$ (this 
corresponds to $n_x = n_y = 0$ in Eqs.~\eqref{overlap5}) has $p_\bk (t)$ 
varying substantially with $t$, and all other values of $\bk$ have $p_\bk (t) 
\approx 1$ for all $t$. Using the expressions for $p_{n_x,n_y}$ in 
Eqs.~\eqref{overlap5} and setting $p_{0,0} = 0.5$, we find that the values of 
$\bar T$ where ${\cal O}_{\pm,\pm} (T) \approx 0.5$ are given by $0.22/16
\approx 0.014$ in the sector $(+,+)$, $0.22/(9 + 4 \kappa^2)$ ($\approx 
0.024$ for $\kappa = 0.1$) in sectors $(+,-)$, and $(-,+)$, and $0.22/4 
= 0.055$ in the sector $(-,-)$. These numbers also provide estimates of the 
values of $\bar T$ where the oscillations around $t=T/2$ are most prominent.

\begin{figure}
\includegraphics[width=0.45\textwidth,height=0.35\textwidth]{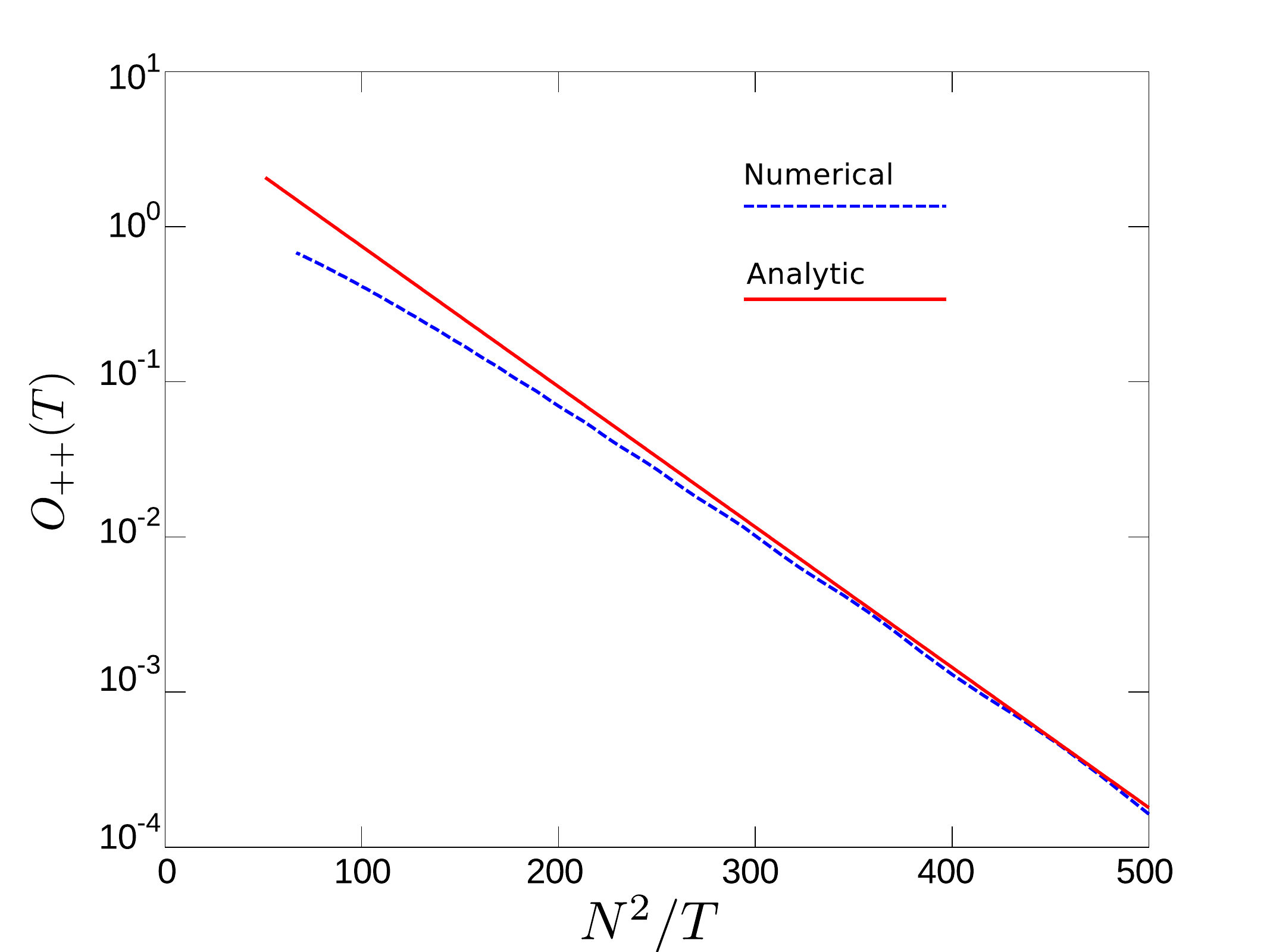}
\caption{The numerical scaling of the final overlap as a function of the 
$N^2/T$ against the predicted behavior in Eq.~\eqref{overlap6}. The system 
size used here is $N_x=N_y=100$ with $J_x=J_y=1$, $\kappa=0.25$ and $J_z = 
3 - (2 t/T)$.}
\label{fig:scaling} \end{figure}

In Fig.~\ref{fig:overlap_kitaev} we show the numerically calculated overlaps for the 4-sectors of the honeycomb model. The figure was calculated using the 2-dimensional equivalent of  Eq.~\eqref{eq:Ut} and clearly illustrates the predicted sectoral dependence of the overlap in quenching from the A phase into the B phase.  We note  however that the $O_{++}$ is the overlap between time-evolved state and the state that would be reached by adiabatic transport of the initial state (that is, the lowest energy state in this sector with the same momentum).  In this case, unlike the blocked sector of the transverse Ising model, the adiabatically transported state is not the lowest energy state in the band, (see Figure \ref{fig:topblockhc}).

In Fig.~\ref{fig:scaling}, we compare the scaling of the final overlap 
${\cal O}_{++} (T)$ obtained numerically versus the scaling predicted in
Eq.~\eqref{overlap6}. We see that the agreement is good at large values of 
$N^2/T$ (where $N^2 = N_x N_y$), but there are some deviations at small 
values of $N^2/T$. The reason for the latter is as follows. In going from 
Eq.~\eqref{overlap5} to \eqref{overlap6} for the log of the overlap, we have 
replaced the sums over $n_x,n_y$ by integrals. This is justifiable only if 
the terms being summed over vary slowly with $n_x,n_y$. However, if 
${\bar T} = \pi^2 JT/N^2$ is large, we can see from Eq.~\eqref{overlap5}
that the terms vary rapidly with $n_x,n_y$, going to zero quickly as 
$n_x,n_y$ increase.

\section{Disorder effects}
\label{sec:disorder}

So far, we have discussed topological blocking and the dependence of 
the quench behavior of overlaps on topological sectors in fermion/spin models 
that preserve translational invariance. However, despite the crucial difference between the un-blocked and blocked sectors (the former is gapped while there exists a gapless spectrum above the sectoral ground state
of the blocked sector) one does  not observe any real qualitative difference in the post quench overlap behavior.  This is because, even in the gapless sector, there is an effective gap to the lowest energy excited state with the same momentum as the ground state. However, a disordered quench will mix these different momenta and thus we can then observe the major characteristic differences between the quenches in the different topological sectors. 

Here, we bring out this qualitative difference by analyzing a disordered version of the transverse Ising system. Our disordered quench protocol, which we numerically implement in the 
one-dimensional case, involves explicitly randomizing in position space the quench term in Eq.~\eqref{eq:hquench} as $h_i(t) = h(t) + V_i $ . Here the $V_i$ are random values from a Gaussian distribution with standard deviation $\sigma$.  We keep this additional term fixed for the duration of the quench. It is important to note that in the presence of disorder the eigenvalues of the  operator $T_z$ are still good quantum numbers even though we cannot directly associate them with discrete momenta.

\begin{figure}
\includegraphics[width=0.45\textwidth,height=0.35\textwidth]{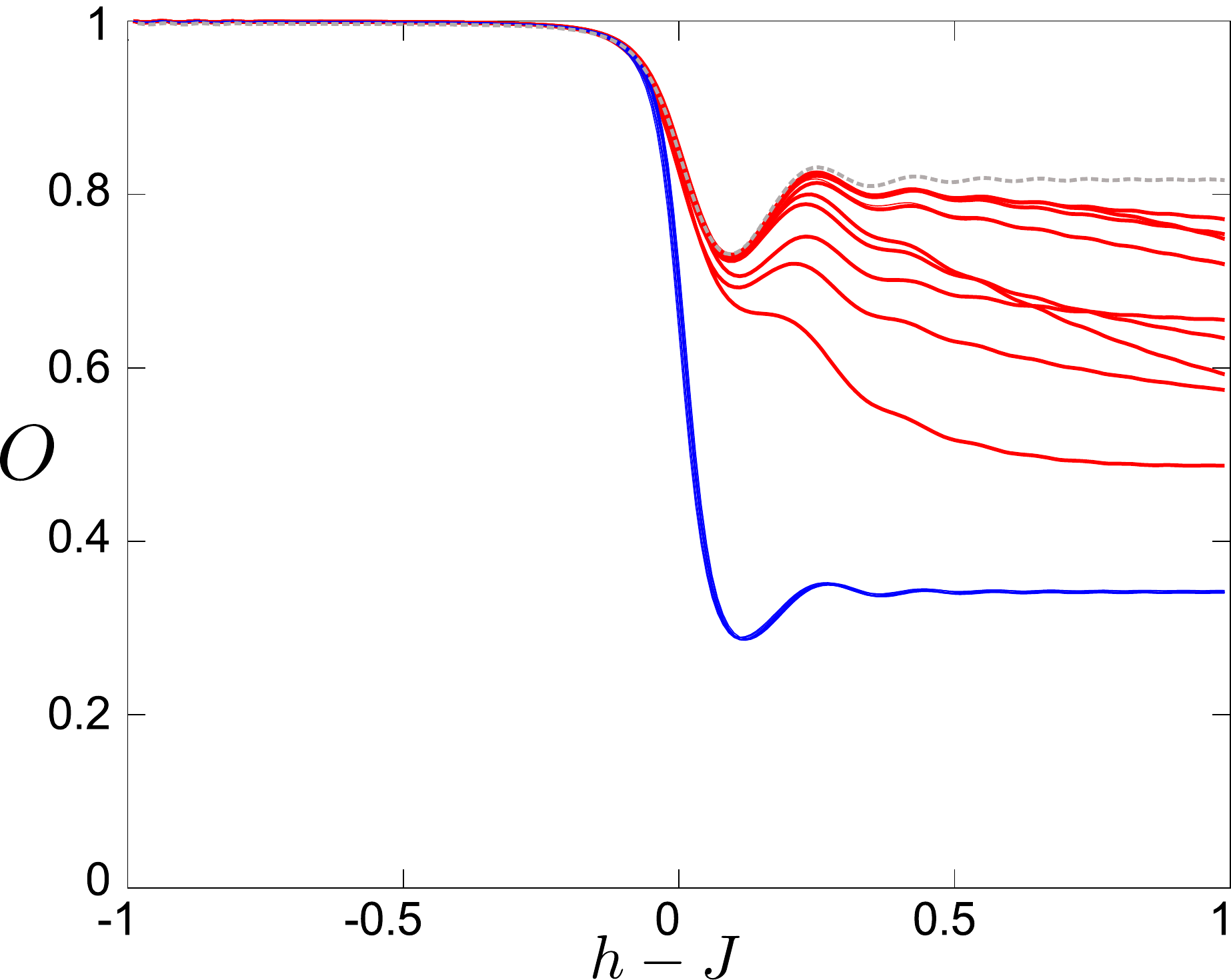}
\caption{Instantaneous ground state overlap for a number of constant in time disorder 
configurations. Disorder has a negligible effect on the overlap profile of the 
unblocked (blue) sector. On the other hand we see that disorder allows the 
blocked sector (red) to disperse within the band inducing a clear instability 
in the overlap profile. The zero disorder values are indicated here by the 
dashed grey line for comparison. In this figure we examined a $N=60$ site 
chain with $J=1$ over a time $T=50$.  The disorder configurations used have standard deviations $\sigma=0.1$. } \label{fig:disorder1} \end{figure}

In Fig.~\ref{fig:disorder1} we show numerical results for overlaps for
multiple quenches with different static disorder potentials. Compared with the 
translationally invariant case (dashed grey line) we see that the behavior of 
the blocked state overlap is characteristically different immediately after 
the quench. This is a generic phenomenon that we observe for all disorder 
configurations and indicates clearly the gapless nature of the sector.

In principle, we could have considered disorder in any local quadratic-fermion operator 
extending over a few lattice sites and our results would have remained robust. This is because $T_z$ commutes with all such operators and thus remains a good quantum number. For example, we could also have randomized the coupling $J$ in position space.  The fact that there are still two separate sectors, is emphasized in Fig.~\ref{fig:disorder1} through the observation that the two overlaps show distinctly different quench behavior. On the other hand it is important to note that in the spin language, perturbations involving, for example, local $\sigma^x$ operators would break the degeneracy of the initial ferromagnetic phase, giving way to a preferred polarization direction. In the fermionic representation we can understand this by noting that these operators break fermion parity and carry with them strings that violate our fermionic notion of local operations. 

Finally, it should be noted that in situations like the one described here, where we have some topological sectors with a gap and some sectors which are gapless and have a low-lying band, we can always expect to have considerable differences between the sectors' post quench behaviors at finite temperature. This is obviously relevant to any experimental setting in which such quenches might be performed.  If the system is kept in contact with a reservoir at temperature $T$ such that $kT$ is much smaller than the gap in the gapped sectors, but considerably larger than the typical energy spacing in the low-lying band in a gapless sector (in the thermodynamic limit, this spacing goes to zero, while the gapped sectors remain gapped). Regardless of the detailed mechanism of energy exchange between the system and the reservoir, one would then expect that in a slow quench starting in one of the blocked sectors, the system will end up in a thermal mixed state involving many of the states in the low-lying band. As a result, it could be observed with a wide range of momenta.  On the other hand, in the gapped sectors, contact with a reservoir at some temperature well below the gap should have very little effect. Of course, we may also imagine that the presence of a reservoir will eventually mix states in different topological sectors. But the characteristic time for such equilibration should be much longer than the characteristic time for mixing states within a single sector because the Hamiltonian for the interaction between system and reservoir should not depend on the non-local quantum numbers  characterizing the sectors. 

\section{Summary and Outlook}
\label{sec:SnO}
In summary, we have explored the notion of topological blocking, which 
depends on ground state degeneracies, and quench dependence on topological 
sectors as concepts that ought to be applicable to most topological systems. 
We have demonstrated these concepts in the context of topological spinless 
fermionic $p$-wave systems (analogous to superconductors), derived from the transverse Ising chain in 
one dimension and the zero vortex sector of the Kitaev honeycomb model in 
two dimensions. Confining ourselves to translationally invariant systems 
and periodic boundary conditions has allowed us to study decoupled pairs 
of momentum modes respecting Bogoliubov-deGennes Hamiltonians. Topological 
sectors and degeneracies have been identified in terms of fermion parity, 
dictated by the occupation numbers of special unpaired momentum modes. In the 
Ising systems, we have illustrated topological blocking in quenching from 
the double degeneracy topological phase to the non-topological phase with a 
unique ground state, and in the Kitaev honeycomb system, from a four-fold 
degenerate Abelian phase to a three-fold degenerate non-Abelian phase. Our 
analytic treatment of quench within different topological sectors has 
involved employing Landau-Zener physics within each momentum sector and has 
been corroborated by numerics. We have found that a sensitive measure of 
quench dependence on topological sectors is the overlap between the 
time-evolved initial ground state within a sector and the sectoral ground 
state of the final Hamiltonian, or more precisely the overlap between the time evolved state and the state that it would evolve to in the adiabatic limit. 
Finally, by numerically incorporating disorder
in our quench protocol, we have shown that quench behavior in different 
topological sectors can be qualitatively very different, particularly if the 
blocked sector can access a gapless spectrum. 

Given that, to the best of our knowledge, this is the first study to 
explicitly address degeneracies by way of topological blocking and 
distinguishing topological sectors via quench dynamics, there are several 
avenues for further investigation. Our analyses of the dynamic behavior of 
overlaps and of other quantities are by no means exhaustive; we hope to 
develop these further. While our quench protocol has involved a linear quench,
several studies have investigated the effect of non-linear 
quenches~\cite{Mondal08,Degrandi08}; 
it would be worthwhile to ask whether these quenches 
can highlight topological aspects better than the linear quench. Starting with 
the original Kibble-Zurek treatments, several works have considered quench 
physics in terms of defects, vortices and vortex loops, and it would be 
interesting to see if these entities have different structures that depend
on the topological sectors. As for the treatment of disorder in the last 
section, our studies are very preliminary. There is scope for an
extensive study bringing out qualitative differences between sectors and 
making connections with other work on disordered quenches e.g. Refs.~\onlinecite{Fine2009,Brandino2012,Kolley2012}.
In the Kitaev honeycomb system, translational symmetry can also be broken by considering the system away from the zero vortex sector; the inclusion of vortices amounts to changing signs on the bonds in a lattice model of 
the $p$-wave superconductor. Such a study could also tie in 
with predicted vortex-nucleation properties (see for example Refs.~\onlinecite{Gils09,Lahtinen2013}) .
 
Turning to other topological systems of interest, the fractional quantum Hall 
systems are a paradigmatic example of topological order, extensively studied for 
their degeneracy properties on the torus. In principle, some of the various 
techniques used to analyze quantum Hall systems can also be employed to study 
quenching in the context here. While Abelian states, such as Laughlin states 
would perhaps be simpler to analyze, non-Abelian states would be of much 
interest in the parallels with chiral superconductors ($\nu = 
5/2$)~\cite{Read00}.  It would also be of great interest to study quenching in systems with symmetry protected topological order\cite{Pollmann2012,Gu2009}, such as topological insulators.
Experimental settings for studying features discussed here 
could also include spin chains, the recently realized topological superconducting 
wires~\cite{Mourik12}, lattice models in cold atomic 
gases~\cite{Greiner2002,Kinoshita2006,Simon2011}, and 
quantum Hall systems.

\section*{Acknowledgments}

We thank A. Chandran, F.  Essler and J. Vala for insightful comments.
G.K. acknowledges the financial support of Science Foundation 
Ireland under the Award 10/IN.1/I3013 and the Alexander von Humbolt Foundation,
D.S. thanks DST, India under Project No. SR/S2/JCB-44/2010, J.K.S acknowledges 
funding from Science Foundation Ireland grants 08/IN.1/I1961 and 12/IA/1697,
and S.V. thanks the Simons Foundation under Grant No.229047 and the National 
Science Foundation under grant DMR 0644022-CAR. We particularly wish to thank
the Aspen Center for Physics where these studies were initiated.

\end{document}